\documentclass[sigplan]{acmart}

\def\BibTeX{{\rm B\kern-.05em{\sc i\kern-.025em b}\kern-.08emT\kern-.1667em\lower.7ex\hbox{E}\kern-.125emX}}
\copyrightyear{2019} 
\acmYear{2019} 
\setcopyright{acmcopyright}
\acmConference[ASPLOS '19]{2019 Architectural Support for Programming Languages and Operating Systems}{April 13--17, 2019}{Providence, RI, USA}
\acmBooktitle{2019 Architectural Support for Programming Languages and Operating Systems (ASPLOS '19), April 13--17, 2019, Providence, RI, USA}
\acmPrice{15.00}
\acmDOI{10.1145/3297858.3304018}
\acmISBN{978-1-4503-6240-5/19/04}

\usepackage{amsmath}
\usepackage{graphicx}
\usepackage[normalem]{ulem}
\usepackage{bm}
\usepackage{color}
\usepackage{braket}
\usepackage{algorithm}
\usepackage{algorithmic}
\usepackage{verbatim}
\usepackage{appendix}
\usepackage{xcolor}
\usepackage{sidecap}
\usepackage{qcircuit}
\usepackage{array}
\usepackage{balance} 

\DeclareCaptionLabelFormat{figure}{Fig.\nobreakspace\thefigure}
\DeclareCaptionLabelFormat{table}{Tab.\nobreakspace\thetable}
\usepackage[compact]{titlesec}

\begin{document}

\title{Optimized Compilation of Aggregated Instructions for Realistic Quantum Computers}

\author{Yunong Shi}
\affiliation{%
  \institution{The University of Chicago}
  }
\email{yunong@uchicago.edu}

\author{Nelson Leung}
\affiliation{%
  \institution{The University of Chicago}
  }
\email{nelsonleung@uchicago.edu}

\author{Pranav Gokhale}
\affiliation{%
  \institution{The University of Chicago}
  }
\email{pranavgokhale@uchicago.edu}

\author{Zane Rossi}
\affiliation{%
  \institution{The University of Chicago}
  }
\email{zmr@uchicago.edu}

\author{David I. Schuster}
\affiliation{%
  \institution{The University of Chicago}
  }
\email{david.schuster@uchicago.edu}

\author{Henry Hoffmann}
\affiliation{%
  \institution{The University of Chicago}
  }
\email{hankhoffmann@cs.uchicago.edu}

\author{Frederic T. Chong}
\affiliation{%
  \institution{The University of Chicago}
  }
\email{chong@cs.uchicago.edu}

\begin{abstract}
Recent developments in engineering and algorithms have made real-world applications in quantum computing
possible in the near future. Existing quantum programming languages and compilers use a quantum assembly language composed of 1- and 2-qubit (quantum bit) gates. Quantum compiler frameworks translate this quantum assembly to electric signals (called control pulses) that implement the specified computation on specific physical devices.
However, there is a mismatch between the operations defined by the 1- and 2-qubit logical ISA and their underlying physical implementation, so the current practice of directly translating logical instructions into control pulses results in inefficient, high-latency programs.
To address this inefficiency, we propose a universal quantum compilation methodology that aggregates multiple logical operations into larger units that manipulate up to 10 qubits at a time.  Our methodology then optimizes these aggregates by (1) finding commutative intermediate operations that result in more efficient schedules and (2) creating custom control pulses optimized for the aggregate (instead of individual 1- and 2-qubit operations).  Compared to the standard gate-based compilation, the proposed approach realizes a deeper vertical integration of high-level quantum software and low-level, physical quantum hardware.  We evaluate our approach on important near-term quantum applications on simulations of superconducting quantum architectures. Our proposed approach provides a mean speedup of $5\times$, with a maximum of $10\times$.  Because latency directly affects the feasibility of quantum computation, our results not only improve performance but also have the potential to enable quantum computation sooner than otherwise possible.

\end{abstract}

\maketitle

\section{Introduction}\label{Introduction}

The past twenty years have seen the world of quantum computing moving closer to solving classically intractable problems \cite{Boixo2016, Farhi2016, Shor1994}. With developments in Noisy Intermediate-Scale Quantum (NISQ) \cite{Preskill2018} devices like IBM's quantum machine with 50 qubits and Google's quantum machine with 72 qubits,  we may soon be able to demonstrate computations not possible on classical supercomputers \cite{Boixo2016, Farhi2016}. Exciting classical-quantum hybrid algorithms tailored for NISQ machines, like Quantum Approximate Optimization Algorithm (QAOA) \cite{Farhi2014} and Variational Quantum Eigensolver (VQE) \cite{Mcclean2016,Peruzzo2014} will power up the first real-world quantum computing applications with scientific and commercial value.

Computation latency is a major challenge for near-term quantum computing.  While all computing systems benefit from reduced latency, in a quantum system the output fidelity decays at least exponentially with latency \cite{NielsenChuang}.  Thus, in near-term quantum computers, reducing latency is not just a minor convenience---latency reduction actually enables new computations on near-term machines by ensuring that the computation finishes before the qubits decohere and produce a useless result. Thus latency reduction is critical to enabling quantum computing applications on near-term NISQ devices.

\begin{figure}[h]
      \centering
      \vspace*{.2in}
\includegraphics[scale=0.640]{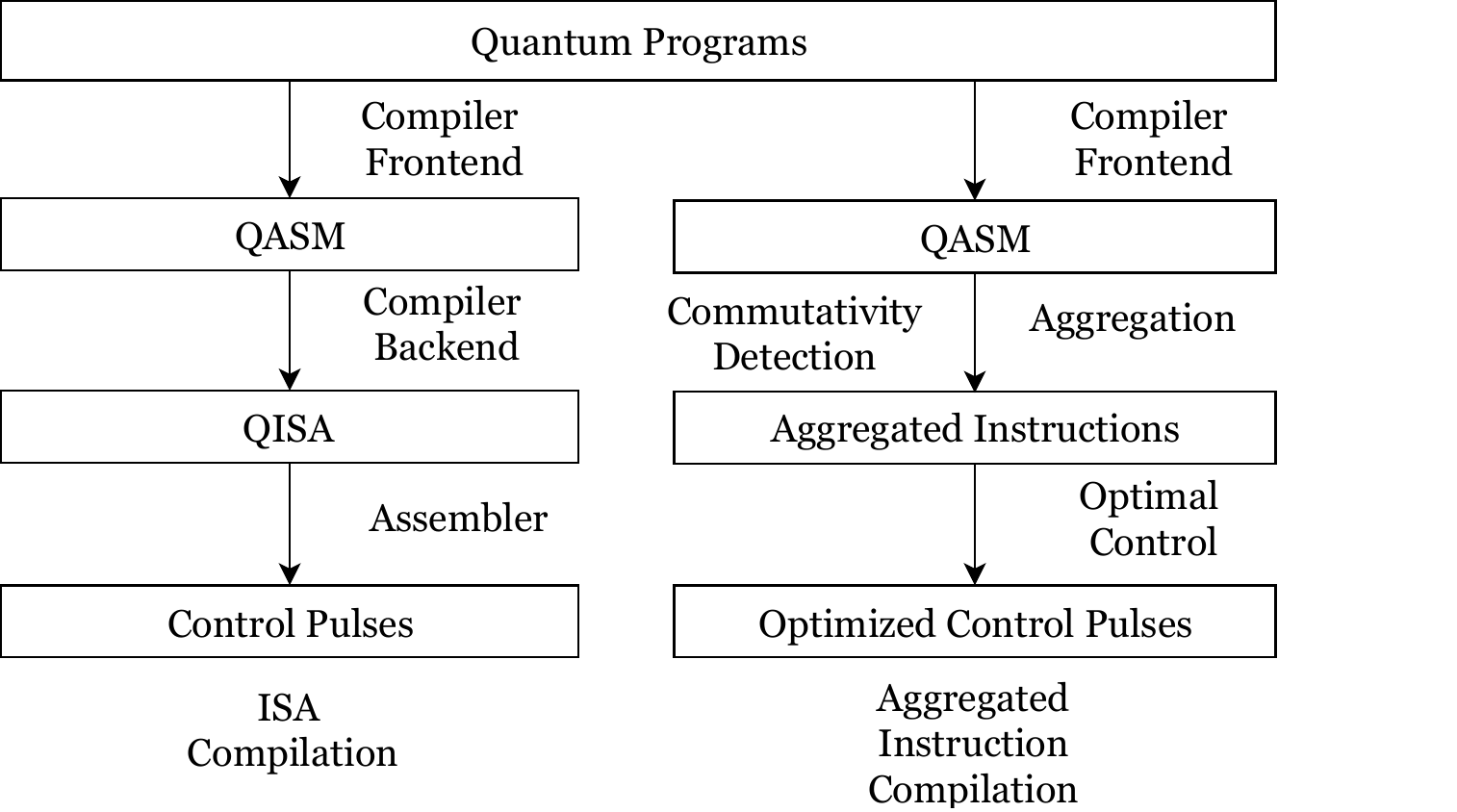} 
\caption{
Comparison of two compilation schemes. Gate-based compilation with ISA abstraction (left) follows a classical compilation approach, but could generate unoptimized quantum operations in the hardware.  Our proposed approach (right) produces highly optimized control pulses.}

\label{simple_flow}
\end{figure}

Unfortunately, existing quantum computing abstractions (which mirror classical computer system stacks, as shown on the left side of Figure \ref{simple_flow}) introduce inefficiencies that greatly impact latency.  In these \emph{gate-based} approaches, programs are compiled into quantum assembly instructions (or \emph{gates}) that specify 1- and 2-qubit operations \cite{Haner2018, prac_ISA, Fu2017}.  This quantum assembly is a virtual ISA which represents a rich set of operations.  These gates must then be translated into \emph{control pulses}---the electrical signals that implement the specified operations on the underlying physical hardware.  Typically though, the underlying hardware implements a different set of operations, and there is a mismatch between the expressive logical gates and the set of instructions that can be efficiently implemented on a real system.  In contrast, physicists have developed a set of techniques---\emph{quantum optimal control}---that ignore abstraction barriers and produce customized control pulses that minimize latency for a particular computation on a physical system \cite{Glaser2015}.  To draw an analogy to classical computer systems, the gate-based compilation approach is similar to the compiler-architecture-microarchitecture stack, while quantum optimal control is similar to customized circuit design.  Quantum optimal control techniques do not scale, however, and are impractical for computations using more than 10 qubits \cite{Nelson2017}, i.e., emerging NISQ systems.

In this paper we propose a quantum compilation technique that optimizes across existing abstraction barriers to greatly reduce latency while still being practical for large numbers of qubits.  Specifically, rather than directly translating 1- and 2-qubit gates to control pulses, our framework aggregates these small gates into larger operations, as illustrated in the right side of Figure \ref{simple_flow}.  Our framework can then manipulate these aggregates in two ways.  First, it finds commutative operations that allow for much more efficient schedules of control pulses.  Second, it uses quantum optimal control on the aggregates to produce a set of control pulses that is optimized for the the underlying physical architecture.  Our technique greatly improves efficiency over the existing gate-based compilation methods while mitigating the scalability problem of quantum optimal control methods.  Because ours is a software-based approach, these results can see practical implementation much faster than experimental approaches for improving physical device latency.  We compare our methodology to standard gate-based compilation on important near-term quantum algorithms and find that our technique produces a mean speedup of $5 \times$ with a maximum speedup of  $10\times$.

We achieve these speedups via two novel techniques:
\begin{itemize}
    \item detecting diagonal unitaries and scheduling commutative instructions to reduce the critical path of computation.
    \item blocking quantum circuits in a way that  scales optimal control beyond 10 qubits without compromising parallelism
\end{itemize}

For quantum computers, achieving these speedups (and thereby reducing latency) is do-or-die: if circuits take too long, the qubits decohere by the end of the computation. By reducing latency 2-10x, our methodology provides an accelerated pathway to running useful quantum algorithms, without needing to wait years for hardware with 2-10x longer qubit lifetimes.

\section{Background}\label{Preliminaries}
This section presents a brief overview of the relevant background on quantum computation and quantum optimal control.
\subsection{Principles of quantum computation}

\begin{SCfigure}
\label{bloch_sphere}
  \centering
  \caption{The Bloch Sphere represents a single qubit. The $\ket{0}$ state is on the North Pole, the $\ket{1}$ state is on the South pole, and superposition states are in between. Single qubit gates correspond to rotations on the Bloch sphere. For instance, the $R_x(\beta)$ gate rotates a qubit by angle $\beta$ about the $x$-axis.\vspace*{.1in}}
  \includegraphics[width=0.2\textwidth]
    {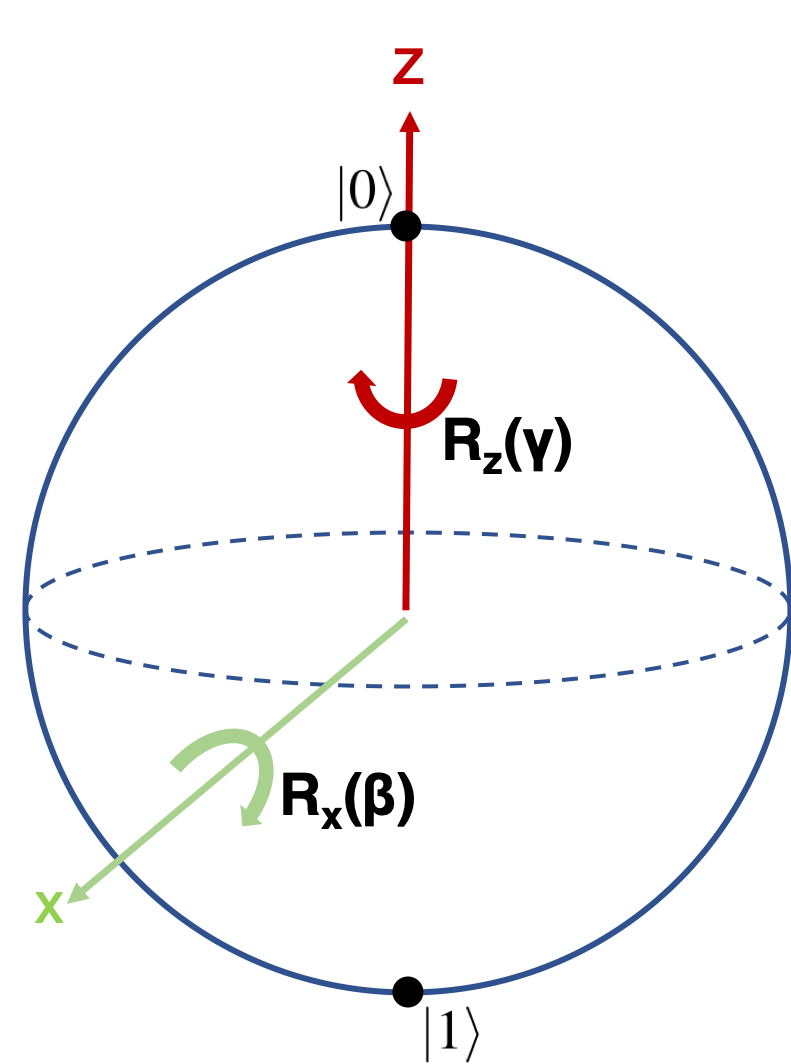}
    \vspace*{.1in}
\end{SCfigure}

The qubit (quantum bit) is the basic element of a quantum computing system. In contrast to classical bits, qubits are capable of living in a superposition of the logical states $\ket{0}$ and $\ket{1}$. The general quantum state of a qubit is represented as $\ket{\psi}=\alpha \ket{0}+\beta\ket{1}$,  where $\alpha, \beta$ are complex coefficients with $|\alpha|^2 + |\beta|^2=1$. When measured in the $0/1$ basis, the quantum state collapses to $\ket{0}$ or $\ket{1}$ with probability of $|\alpha|^2$ and $|\beta|^2$, respectively. It is helpful to visualize a qubit as a point on a 3D sphere called the Bloch sphere \cite{bloch, NielsenChuang}, as depicted in Figure \ref{bloch_sphere}. Qubits can be realized on different Quantum Information Processing (QIP) platforms, including superconducting circuits \cite{Devoret1169}, ion traps \cite{Lekitsche1601540}, and quantum dots systems \cite{quantumdot}. 

The number of quantum logical states grows exponentially with the number of qubits in a quantum system. For example, a system with 3 qubits can live in the superposition of 8 logical states: $\ket{000}$, $\ket{001}$, $\ket{010}$, ..., $\ket{111}$. This property sets the foundation of potential quantum speedup over classical computation---an exponential number of correlated logical states can be stored and processed simultaneously by a quantum system with a linear number of qubits.

\subsection{Quantum gates}\label{lgate}
In the process of quantum compilation, quantum algorithms are first decomposed into a set of universal 1- and 2-qubit discrete quantum operations called logical quantum gates. All gates are represented in matrix form as unitary matrices. 1-qubit gates correspond to rotations along a particular axis on the Bloch sphere. In the standard ISA for quantum computation, the 1-qubit gate set includes rotations along the x-, y-, z-axes of the Bloch sphere, $i.e.$ $R_x$, $R_y$, $R_z$ gate. It also includes the Hadamard gate, which corresponds to rotation about the diagonal x+z axis. An example of a 2-qubit logical gate is the Controlled-NOT (CNOT) gate, which flips the state of the target qubit iff the control qubit is $\ket{1}$. For example, the CNOT gate sends $\ket{10}$ to $\ket{11}$, sends $\ket{11}$ to $\ket{10}$, and preserves the other logical states. 

Because it is typically not obvious how to implement the CNOT gate directly on a physical platform, a CNOT gate is further decomposed into physical gates in standard gate-based compilation. Appendix \ref{app1} provides a description of 2-qubit physical gates on different quantum platforms. For the benchmarks we present in this paper (Section \ref{experiment}), we focus on superconducting architectures with the iSWAP physical gate because it is easy to implement and its optimized compilation is relatively unexplored.

\subsection{Quantum control}
Quantum computing systems can be continuously driven by external physical operations to any state in the space spanned by the logical states. The physical operations, called control fields, are specific to the underlying system, with control fields and system characteristics controlling a unique and time-dependent quantity called the Hamiltonian. The Hamiltonian determines the evolution path of the quantum states.  For example, in superconducting systems, we can drive a qubit to rotate continuously on the Bloch sphere by applying microwave electrical signals \cite{jerry}. By varying the intensity of the microwave signal, we can control the speed of the qubit's rotation. The ability to engineer the system Hamiltonian in real-time allows us to direct the qubits to the quantum state of interest through precise control of related control fields. Thus, quantum computing is achieved by constructing a quantum system in which the Hamiltonian evolves in a way that aligns with a computational task, yielding the desired result with high probability upon final measurement of the qubit system. In general, the path to a final quantum state is not unique and finding the optimal evolution path is an open problem \cite{Nelson2017, CISC2007, Glaser2015}.

In the context of quantum control, quantum gates can be regarded as a set of pre-programmed control fields performed on the quantum system.

\subsection{The mismatch between gates and control}
The coarse-grained abstraction of quantum gates can complicate the  continuous evolution of the underlying quantum states, meaning that the pre-programmed control fields might not lead to the  optimal evolution path of the quantum system. We consider two simple examples to illustrate this mismatch.

\begin{itemize}
    \item In the first example, consider the gate sequence of a CNOT gate followed by a $R_z$ gate. In standard gate-based compilation, these two logical gates will be further decomposed into physical gates and be executed sequentially. However, on superconducting platforms, the control fields that implement the two gates can be applied simultaneously. Hence, in this case, the gate model is suboptimal as it serializes the circuit and thus increases the circuit latency.
    \item 
As another example, consider the traditional ISA decomposition of the SWAP operation into three Controlled-NOT (CNOT) operations, as realized in the circuit below. This decomposition is equivalent to the implementation of in-place memory SWAPs with three alternating XORs in classical computation. For systems like quantum dots \cite{quantumdot}, the SWAP operation is directly supported by applying particular constant control fields for a certain period of time. In this case, decomposing a SWAP into three CNOTs introduces substantial overhead.

 \begin{figure}[h]
      \centering
\hspace*{-0.1cm}\includegraphics[scale=0.25]{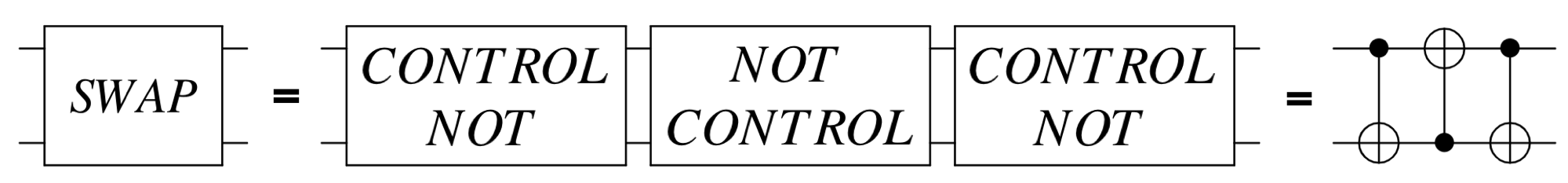} 
\label{swap_decomposition}
\end{figure}

 \begin{figure}[t]
      \centering
\hspace*{-0.4cm}\includegraphics[scale=0.320]{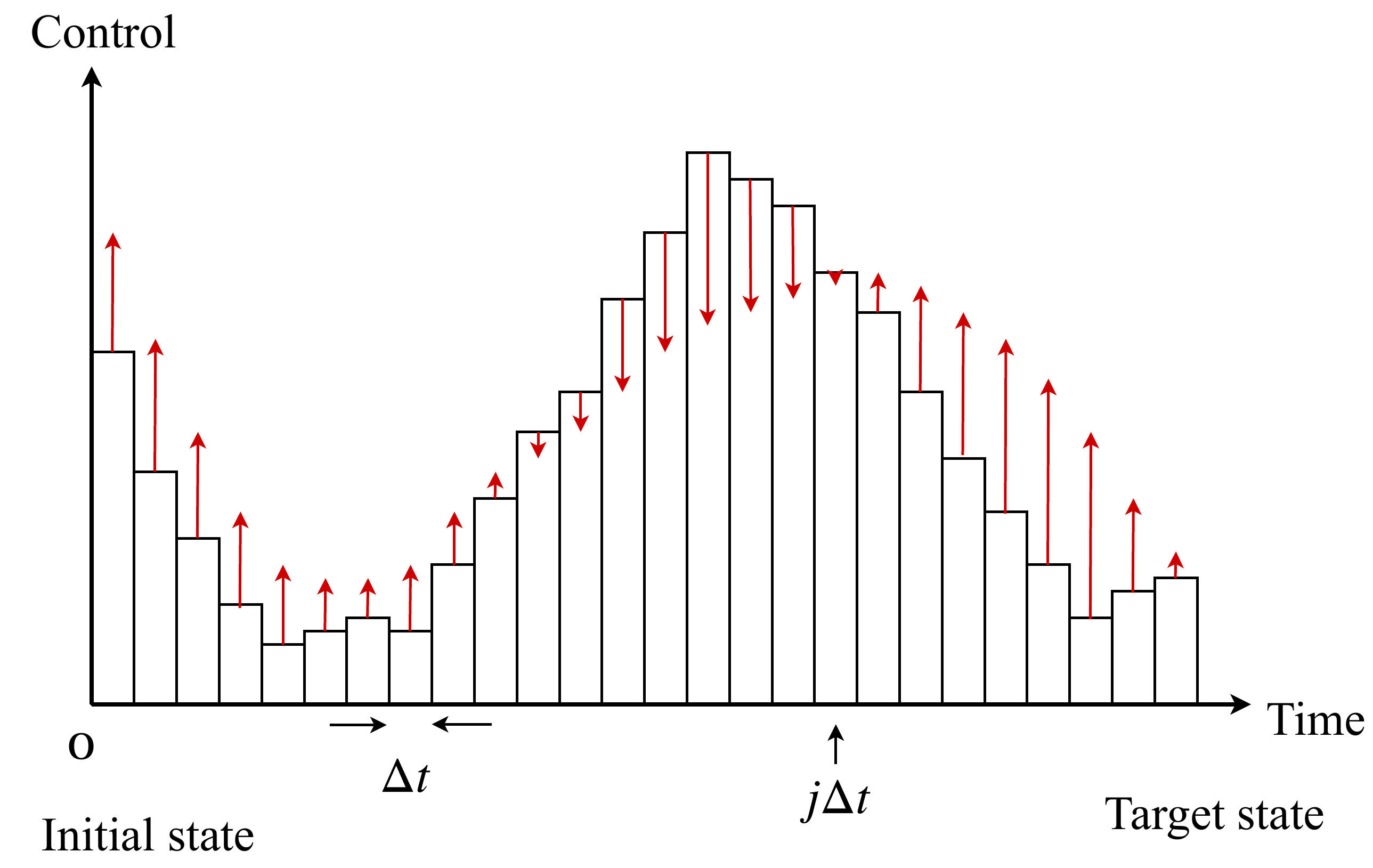} 
\caption{Quantum optimal control based on gradient descent, for a simplified single-pulse-type example. The black bars indicate the current iteration's proposed sequence of control pulse amplitudes by time interval, $\mu(j)$. The red arrows indicate the gradient of the output fidelity with respect to each $\mu(j)$.  Thus, at the next iteration, each amplitude should be updated to $\mu(j) + \epsilon\frac{\partial L}{\partial \mu(j)}$, where $L$ is the targeted loss function and $\epsilon$ is the adaptive step size.}
\label{grape}
\end{figure}
\end{itemize}
In experimental physics settings, equivalences from simple gate sequences to control pulses can be hand optimized \cite{schuch2003}. However, when circuits become larger and more complicated, this kind of hand optimization become less efficient and the standard decomposition becomes less favorable,  motivating a shift toward numerical optimization methods that are not limited by the ISA abstraction.

\subsection{Quantum optimal control}
 Quantum optimal control algorithms find the optimal evolution path from a starting quantum state to a final quantum state, typically by performing gradient descent methods, such as the GRadient Ascent Pulse Engineering (GRAPE) \cite{grape1, grape2} algorithm. For a quantum system with a set of external control fields $u_1,\ldots,u_M$ that can be tuned in real-time, optimal control minimizes deviations from a target state by adjusting each control field $u$. In GRAPE, at every iteration the gradient of the target loss function (usually fidelity) with respect to a control field $\mu_k$ at time step $j$ in the evolution can be explicitly calculated by solving Schr\"{o}dinger's equation. The algorithm will update the control field $\mu_k(j)$ in the direction of the gradient with adaptive step size $\epsilon$ \cite{grape1, grape2, Nelson2017} (Figure \ref{grape}). With enough iterations, the converged control pulses are expected to drive the system from the initial state to the final state along an optimized path.

 Gradient methods' running time and memory use grow exponentially with the size of the quantum system. In our work, we are able to numerically optimize quantum systems of up to 10 qubits with the GPU accelerated optimal control unit \cite{Nelson2017}.
\section{Compilation methodology}\label{methodology}

In this section, we demonstrate by example the advantage of our approach over standard gate-based compilation. Next we present our compilation methodology and introduce its end-to-end tool flow, including the frontend, backend, the optimal control unit, and verification procedure. In Section \ref{calgorithm}, we will detail the instruction aggregation algorithms.

\subsection{An example of QAOA circuit}\label{qaoa}
\begin{figure}[thpb]
      \centering
\includegraphics[scale=0.25]{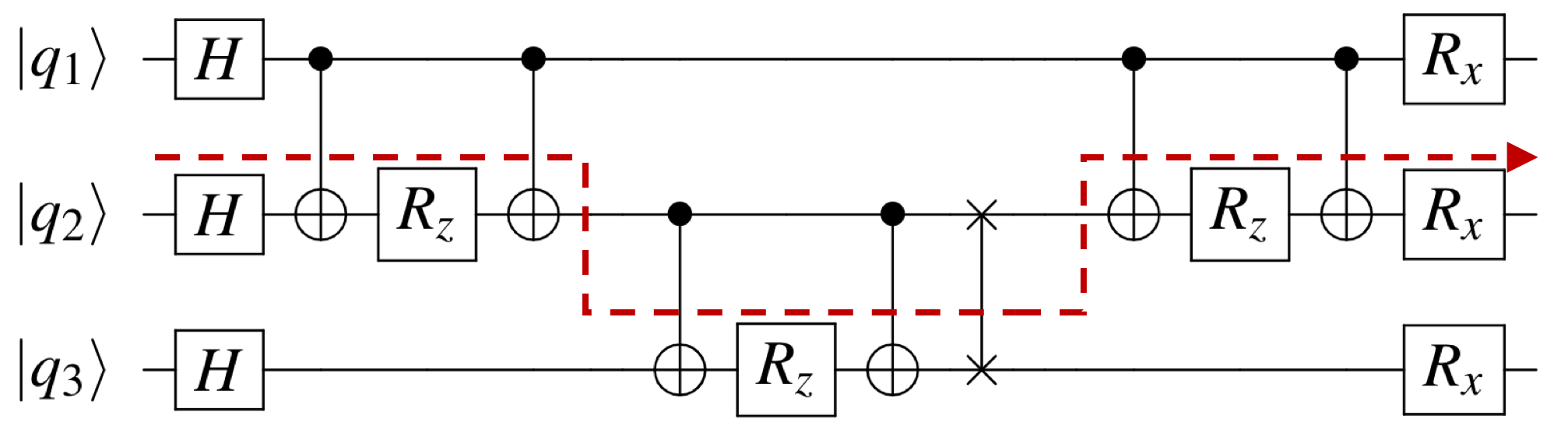} 
\newline
\hspace*{0.3cm}(a)
\includegraphics[scale=0.25]{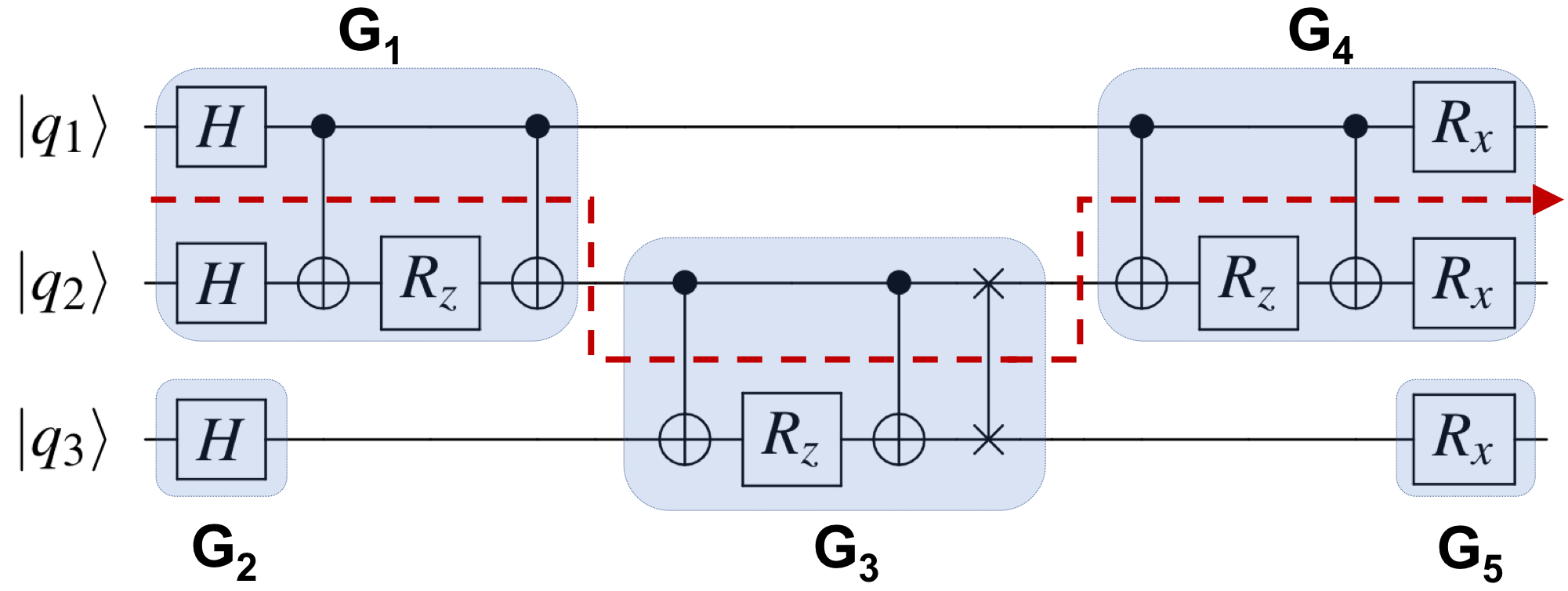}
\vspace*{0.25cm}
\hspace*{2.4cm}(b)
\newline
\newline
\includegraphics[scale=0.46]{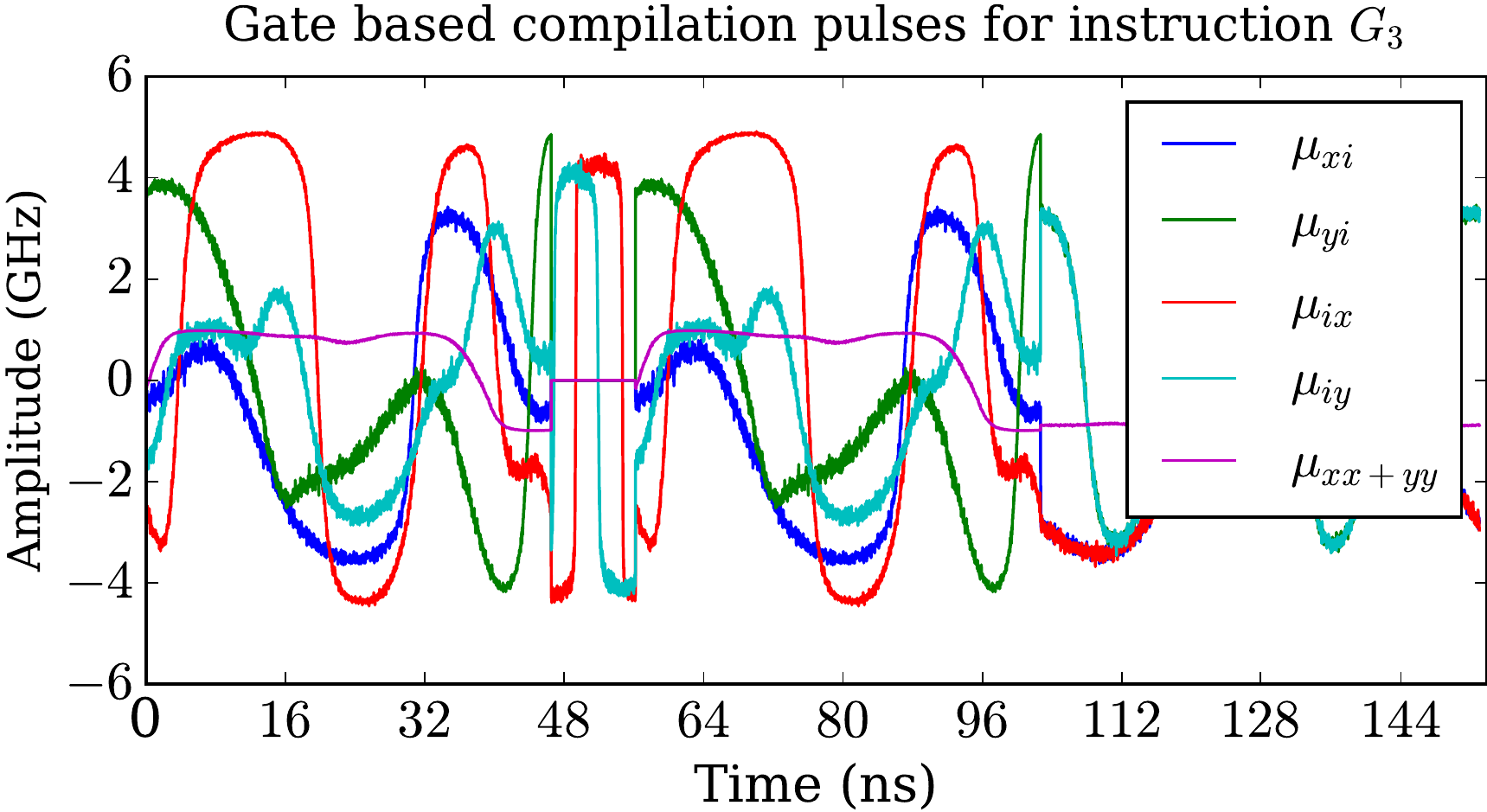}
\vspace*{0.25cm}
\hspace*{2.4cm}(c)
\newline
\newline
\includegraphics[scale=0.46]{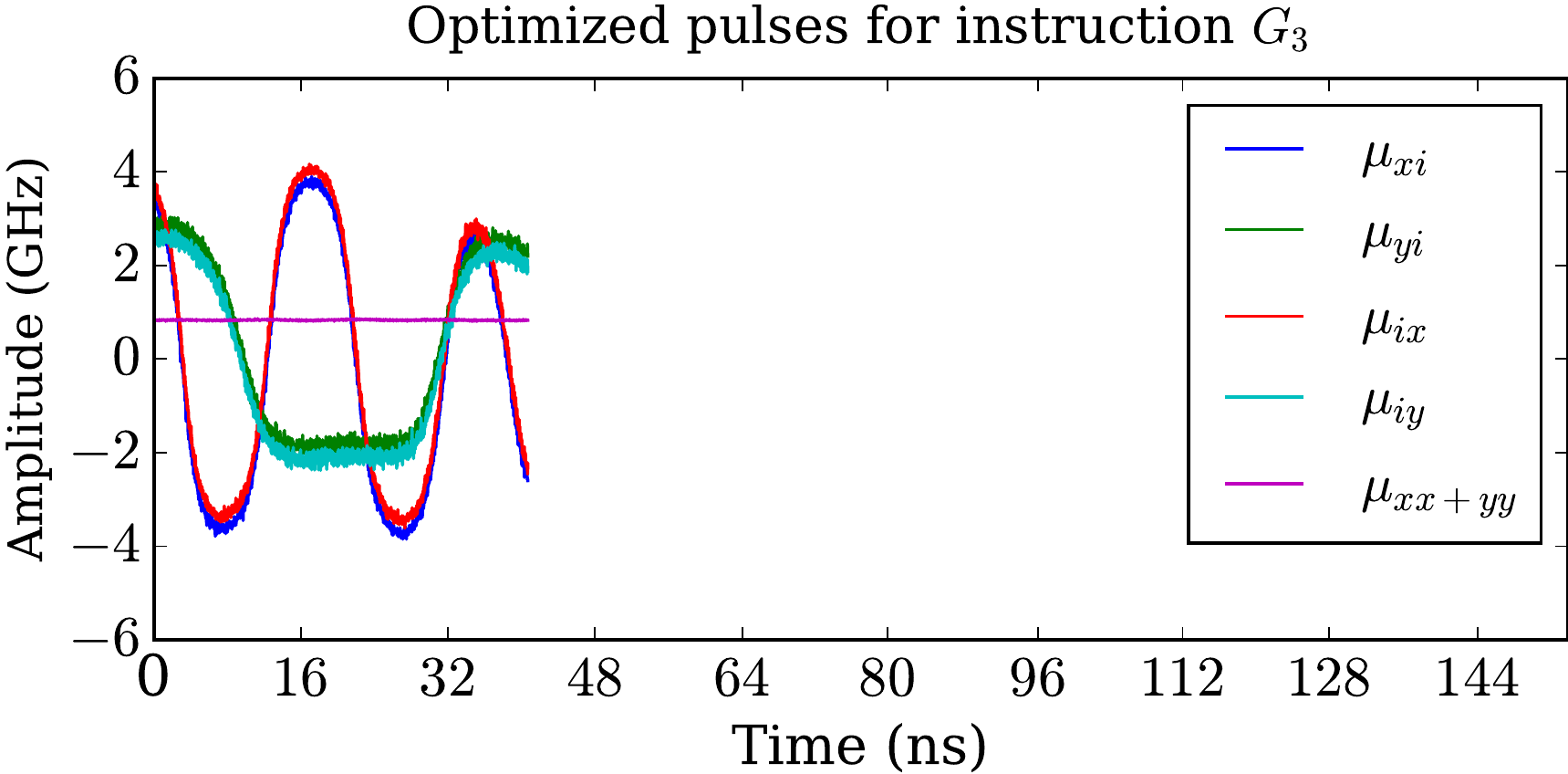}
\hspace*{0.7cm}(d)
\caption{Example of a QAOA circuit demonstrating the difference between gate-based compilation and our compilation methodology. (a)  Standard circuit (red arrow indicates the critical path). (b) Circuit with aggregated instructions. (c) Standard compilation pulses for $G_3$. (d) Aggregated compilation pulses for $G_3$. Each line represents the intensity of a control field. The pulse sequence in (d) is much shorter in duration and easier to implement than that of (c).}
\label{demo_qaoa}
\end{figure}

Figure \ref{demo_qaoa} (a) shows a quantum circuit that solves the MAXCUT problem for a triangle.\footnote{Specifically, the circuit implements the QAOA  \cite{Farhi2014}, one of the most promising near-term quantum algorithms, with angle parameters $\gamma$ and $\beta$ determined by variational methods \cite{Mcclean2016} and set to $5.67$ and $1.26$.} The circuit is decomposed into a standard gate set.
This circuit (or variants of it up to single qubit gates) can be reproduced by most quantum software platforms, including ScaffCC \cite{ScaffCC}, QISKit \cite{openqasm} and Pyquil \cite{pyquil}. We generate this circuit using ScaffCC. To keep our example small and realistic, we assume 1D nearest neighbor qubit connectivity and a underlying superconducting architecture. A SWAP gate is inserted to satisfy the qubit connectivity constraint. We choose to set the 1-qubit control field limit 5$\times$ the 2-qubit control field limit as a representative of real experimental settings \cite{jerry}. The total execution time using gate-based compilation in Figure \ref{demo_qaoa} (a) is found by adding up the pulse time of each individual gate on the critical path of the circuit: $6T(CNOT) + T(SWAP) + T(H) + 3T(R_z) + T(R_x) = 381.9 \text{ns}$ using the numbers in Table \ref{table1}.

\begin{table}[t]
\renewcommand{\arraystretch}{1.3}
\small

\centering
\begin{tabular}{|c|c|c|c|c|c|c}
\hline      
Gate & CNOT& SWAP & H & $R_z(\gamma)$ & $R_x(\beta)$\\
\hline
Time (ns) & 47.1 & 50.1 & 13.7 & 9.8 & 6.1\\
\hline
Gate & $\;\; G_1 \;\;$ & $\;\; G_2\;\; $& $\;\; G_3 \;\;$ & $\;\; G_4 \;\; $ & $\;\; G_5\;\;$\\
\hline
Time (ns)& 54.9 \quad &  13.7\quad& 42.0\quad & 31.4 \quad& 6.1\quad\\
\hline
\end{tabular}
\newline
\vspace{0.15cm}
\caption{Instruction execution time for QAOA circuit in Figure \ref{demo_qaoa} (a). The pulse time for each gate in this table is optimized by an optimal control unit (see section \ref{ocu}). For the SWAP gate, we don't use the standard 3 alternating CNOTs implementation but optimize it individually.}
\label{table1}
\end{table}

\begin{figure*}[thpb]
\centering
     \includegraphics[scale=0.42]{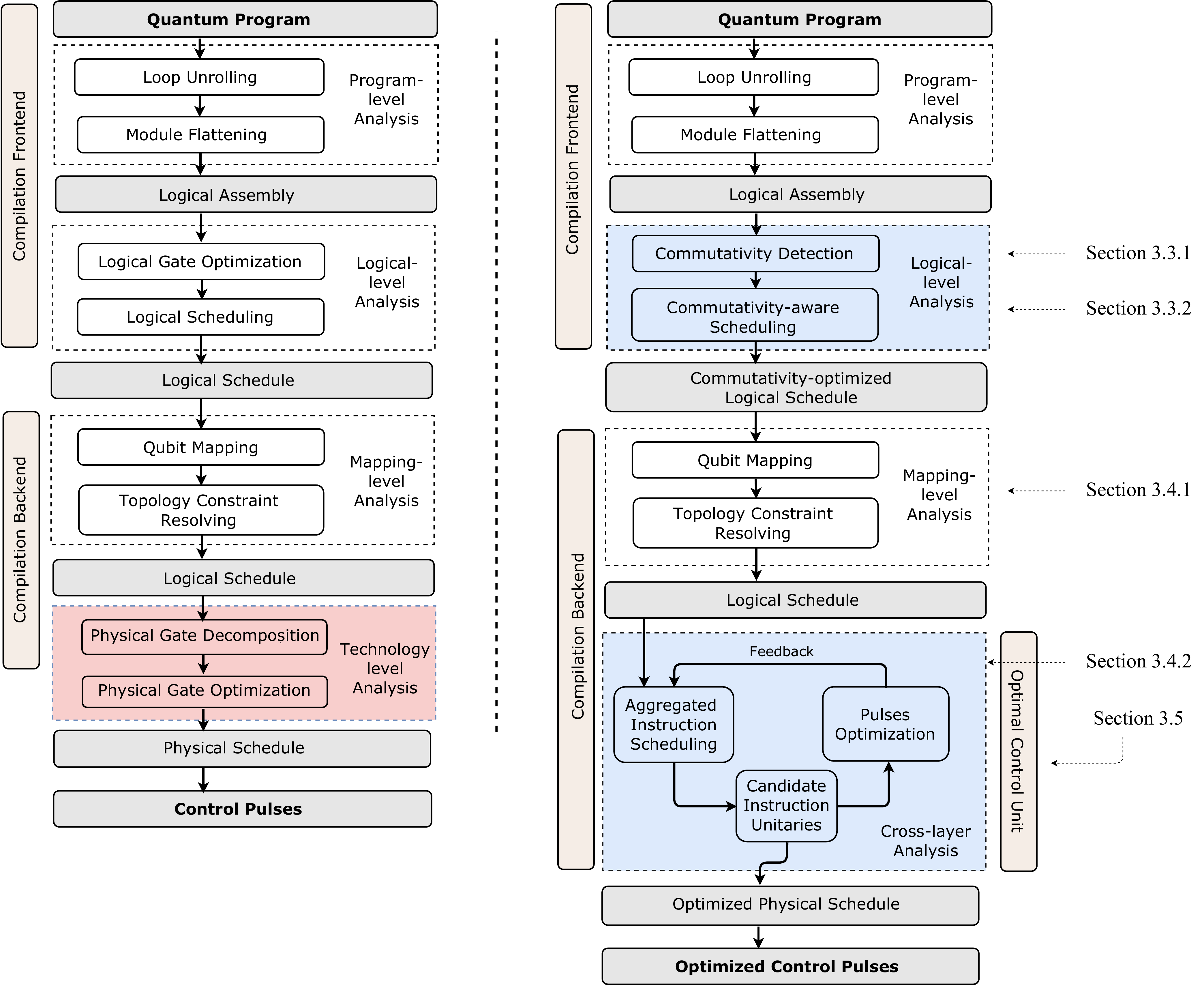}
     \newline
      \caption{ The comparison between standard gate-based compilation (left) and our compilation approach (right). The key differences are highlighted by the colored areas. In the first blue box, our compiler detects potential commutativity, which opens up  opportunities for much more efficient scheduling. Then our logical scheduling takes advantage of commutativity for better parallelization. In the second blue box, by iterating with the optimal control unit, the instruction aggregation procedure breaks the well-encapsulated abstraction of 1- and 2-qubit logical gates and eliminates the physical gate layer (red box) that encodes only coarse-grained hardware information. }
      \label{work-flow}
   \end{figure*}
  
\begin{figure*}[thpb]
      \centering
 \hspace*{-0.3cm}\includegraphics[width=1.04\textwidth]{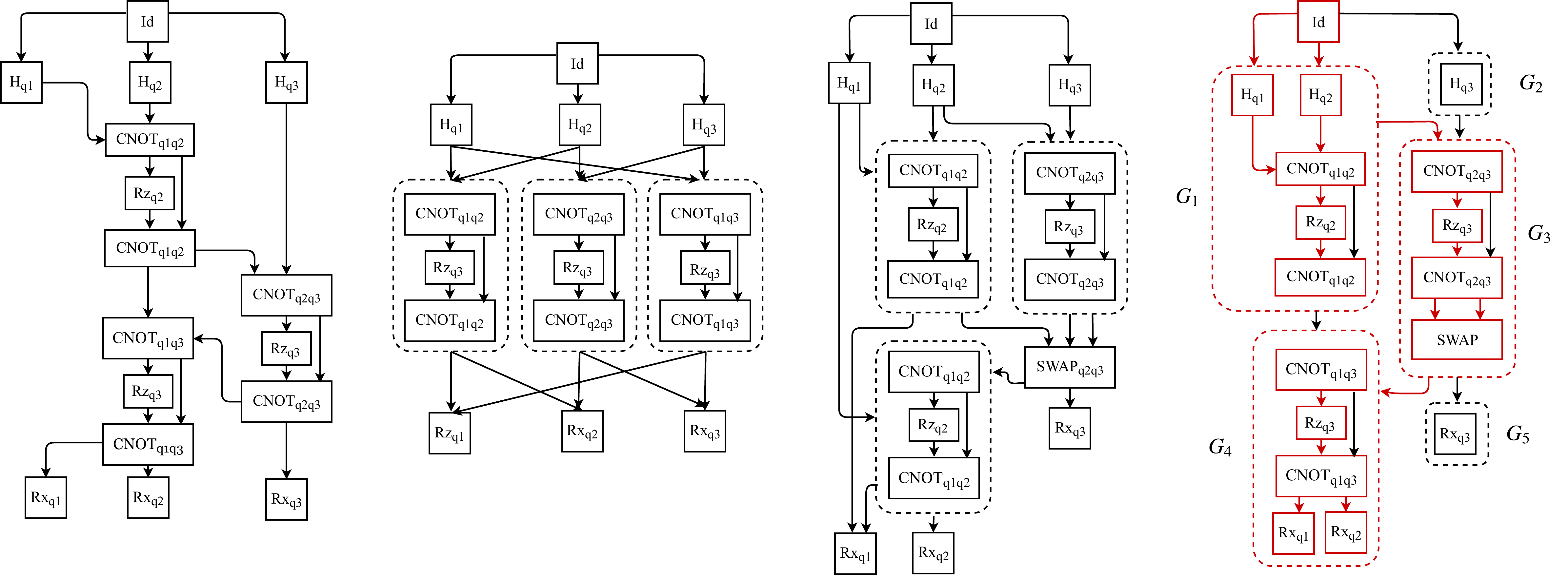}
 \newline
 \newline
 \hspace*{-0.6cm}(a) Module flattening \hspace*{1.2cm} (b) Commutativity detection  \hspace*{0.6cm}     (c) Scheduling and mapping            \hspace*{0.75cm} (d) Gate aggregation
 \vspace*{0.4cm}
   \caption{The evolution of GDG for the circuit in Figure \ref{demo_qaoa}. In the compiler frontend, GDG in (a) is constructed for the flattened quantum program. By detecting commutative CNOT-Rz-CNOT instructions, the compiler transforms the GDG in (a) to GDG in (b) for more scheduling flexibility. Then, after scheduling and mapping, GDG has SWAP gates inserted and becomes GDG in (c). Finally, after the final aggregation,  we arrive at the final GDG in (d), which is optimized both for parallelism and pulses generation. Each path in GDGs represents a qubit.  The qubit name for each path is omitted in the figure for cleanness. The red paths in part (d) are the final critical paths.} 
      \label{gdg}
   \end{figure*}

In contrast, our compiler automatically generates the aggregated instruction set $G_1-G_5$ as indicated in Figure \ref{demo_qaoa} (b), and uses optimal control to produce minimal latency pulses for each. The pulse time for the circuit has critical path: $T(G_1)+T(G_3)+T(G_4) = 128.3 \text{ns}$.
In this example, our proposed aggregated instruction compilation reduces the pulse duration by about $2.97\times$ compared to standard gate-based compilation methods. Figure \ref{demo_qaoa} (c) and (d) compare the pulses for $G_3$ generated by gate-based compilation and generated by the optimal control unit.

\subsection{Methodology overview}

   Figure \ref{work-flow} illustrates the key innovations in our proposed compilation scheme compared to standard gate-based compilation. Both approaches take a quantum program as input and proceed through a series of transformations to produce the control pulses that implement the computation on the physical qubits. In the traditional gate-based approach, the compiler first produces flattened quantum assembly codes, then generates a schedule of the logical instructions in the assembly codes.   This schedule is later turned into a schedule of physical instructions by decomposing the logical instructions into physical instructions, which are converted into control pulses. We note that in the traditional gate-based approach, the physical properties of the underlying hardware are "localized" in each physical instruction. Compared to the traditional approach, our compilation process first converts assembly codes to a logical schedule that explores more commutativity by aggregating highly commutative instructions. Unlike traditional logical scheduling, our compiler aggregate highly commutative intermediate instructions in the assembly codes and generates a much more efficient logical schedule by re-arranging the new instructions. The logical schedule is then converted to a physical schedule after qubit mapping and SWAP gate insertion. At this point the compiler aggregates the final instructions and applies optimal control to the aggregated instructions.  The goal is to find the optimal aggregation that produces the lowest-latency control pulses for the specified computation while considering aggregations that are small enough to be processed by the quantum optimal control unit. Output is an optimized physical schedule along with the corresponding optimized control pulses.
   
\subsection{\label{sec:frontend}Compilation frontend}
The compiler frontend accepts quantum programs from the user, lowering high-level descriptions of quantum algorithms
to a logical assembly that retains gate dependence relations. The compiler frontend performs program level analysis and preliminary logical level optimization, including loop unrolling, module flattening, commutativity detection, and logical level scheduling. The logical assembly output from the compiler frontend can be abstracted as a gate dependence graph (GDG) for each program.
\subsubsection*{Quantum GDG:} 
  The main difference between a quantum GDG and a classical program dependence graph (PDG) is that quantum commutation rules apply in quantum GDG. More specifically, in a quantum GDG, consecutive commuting gates do not have parent-child relations \cite{Giacomo2017} and can be scheduled in any order. Important commutation relations are depicted in Table~\ref{tab:commutation_relations}.
  
\begin{table}[]
    \centering
\begin{tabular}{ c | c }
\Qcircuit @C=.2em @R=0em @!R {
& \gate{U} & \qw & \qw & & & \\
& \qw & \gate{V} & \qw & & } \raisebox{-0.8em}{$=$} \Qcircuit @C=.2em @R=0em @!R { & & & \qw & \qw & \gate{U} & \qw \\
& & & \qw & \gate{V} & \qw & \qw } &
\Qcircuit @C=.2em @R=0em @!R {
& \gate{R_z} & \ctrl{1} & \qw & \\
& \qw & \targ & \qw & & &} \raisebox{-0.8em}{$=$} \Qcircuit @C=.2em @R=0em @!R {
& & & \ctrl{1} & \gate{R_z} & \qw \\
& & & \targ & \qw & \qw & \push{\rule[1em]{0em}{0.5em}}
}
\\ \hline

\Qcircuit @C=.2em @R=0.5em @!R {
& \ctrl{1} & \qw & \qw & & & \qw & \qw & \ctrl{1} & \qw \\
& \targ & \targ & \qw & \push{\rule{.3em}{0em}=\rule{.3em}{0em}} & & \qw & \targ & \targ & \qw \\
& \qw & \ctrl{-1} & \qw & & & \qw & \ctrl{-1} & \qw & \qw
} &
\Qcircuit @C=.2em @R=0em {
& \\
& \gate{\begin{smallmatrix} u_{1} & 0 \\ 0 & u_{2} \end{smallmatrix}} & \gate{\begin{smallmatrix} v_{1} & 0 \\ 0 & v_{2} \end{smallmatrix}} & \qw & \push{\rule{.3em}{0em}=\rule{.3em}{0em}} & \qw & \gate{\begin{smallmatrix} v_{1} & 0 \\ 0 & v_{2} \end{smallmatrix}} & \gate{\begin{smallmatrix} u_{1} & 0 \\ 0 & u_{2} \end{smallmatrix}} & \qw & \push{\rule[-3em]{0em}{0em}}\\
} 
\end{tabular}
\caption{Examples of gate commutation relations. Clockwise from top-left: gates acting on different qubits commute, controls commute with Z-axis rotations, gates with diagonal matrices commute, and CNOTs with disjoint controls commute.}
    \label{tab:commutation_relations}
\end{table}
 
 In our compiler frontend, commutation relations between two gates $A$, $B$ are resolved by explicitly checking the equality of unitary operators $\hat{A}\hat{B}$ and $\hat{B}\hat{A}$.
 
Figure \ref{gdg} shows the GDG of the QAOA circuit in Figure \ref{demo_qaoa} at different compilation stages. We insert an identity instruction as a virtual root for every GDG to connect instructions at depth 0. Because this virtual root is the identity instruction, it does not interfere with the computational result or latency. In our GDG, each path is labelled by a corresponding qubit name.

\subsubsection{Commutativity detection:} \label{ps_ai}
Prior to commutativity detection, every consecutively scheduled pair of gates has a parent-child dependence. However, if a pair of gates commutes, then their relationship is a false dependence and the gates can be scheduled in either order. Our compilation technique relies heavily on the flexible scheduling of gates, so detecting commutativity and removing false dependencies in the GDG is critical for the rest of the compilation process.

 In many near-term quantum applications, it is common for instructions \emph{within} an instruction block to not commute, but for the full instruction blocks to commute with each other \cite{Farhi2014, qchemistry}. As an example, in Figure \ref{demo_qaoa}, the CNOT-Rz-CNOT structures commute with each other (these structures correspond to diagonal unitaries), but each CNOT and Rz in these structures does not commute. Thus in the GDG  in Figure \ref{gdg} (b), after contracting the consecutive CNOT, Rz, CNOT instructions, the compiler is able to schedule new commuting CNOT-Rz-CNOT instructions in any order, while in the GDG in Figure \ref{gdg} (a), scheduling options are limited. This observation opens up opportunities for more efficient scheduling. In our design, the commutativity detection step achieves the goal of forming a highly commutative instruction set for the input quantum circuit. We detail the algorithm for commutativity detection in Section \ref{calgorithm}.

\subsubsection{Commutativity-aware Logical Scheduling (CLS):}\label{cschedule}
\begin{figure}[t]
      \centering
\includegraphics[scale=0.3310]{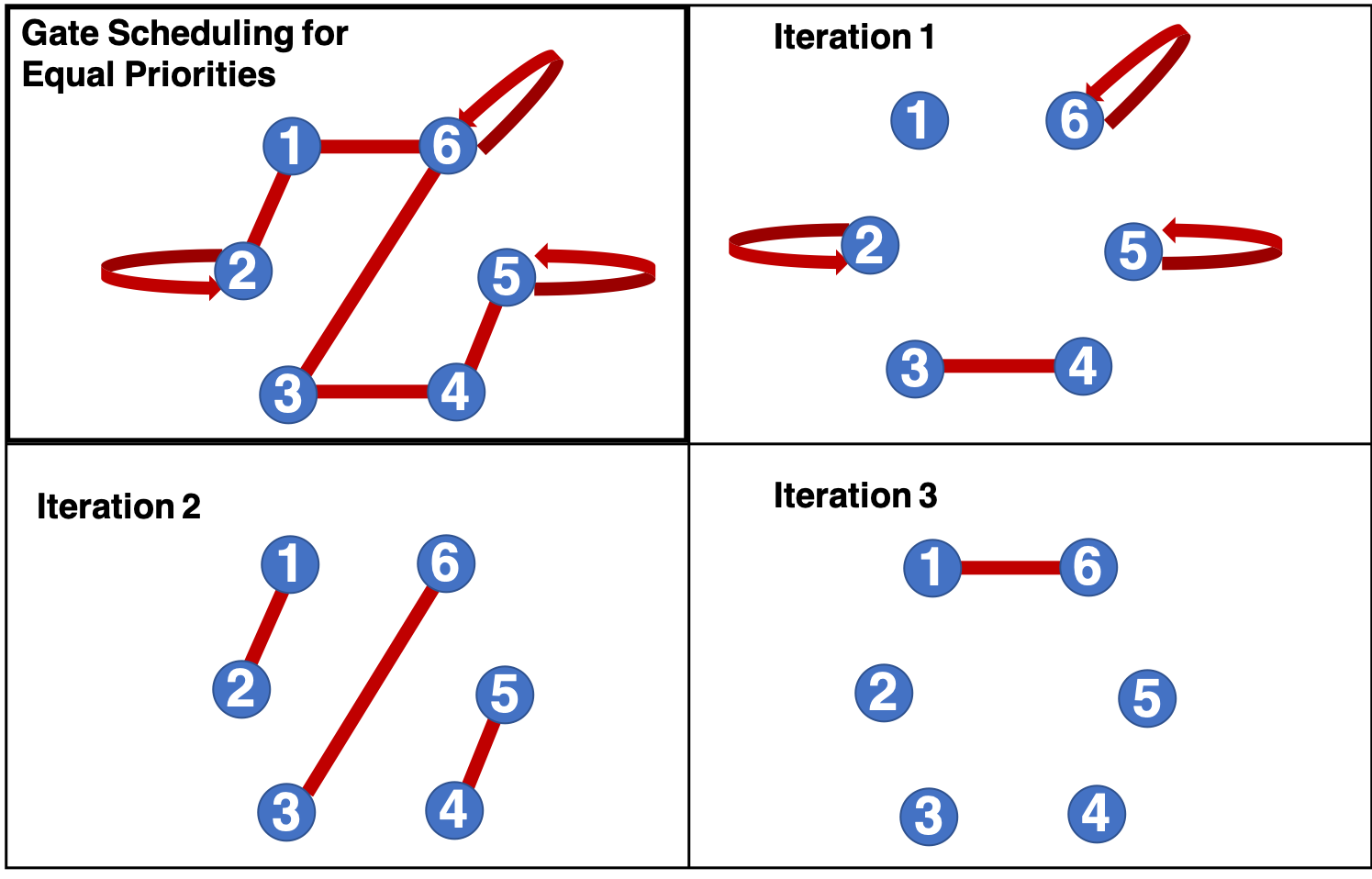}
\caption{A computational graph with six qubits, all instructions have the same latency. The scheduler finds a maximal matching of non-adjacent edges and schedules them. The subsequent round repeats this process on the subgraph of remaining edges.}
\label{schedule}
\end{figure}

CLS uses commutativity, either detected from the last step or inherited from the original circuit, to extract more parallelism. With our GDG construction, it's natural to define the commutation group data structure on qubits. Each qubit maintains a list of commutation groups that, on that qubit, all the consecutive and commutative gates are in the same group. Two gates commute iff they are in the same commutation group on all the common qubits they share. This data structure facilitates more flexible scheduling and more optimization. For example, the two CNOTs in a CNOT-Rz-CNOT structure are in the same commutation group on the control qubit, but in different commutation groups on the target qubit. Then, with the commutation group data structure, we can correctly identify that any Rz gates on their control qubit can travel through these two CNOTs  even though these two CNOTs do not commute. Our CLS iterates the commutation groups on qubits in circuits. At each iteration, the CLS draws candidate gates to schedule from the first non-empty commutation groups on qubits, and schedules greedily. At each step, the candidate gates form a computational graph $G_c$ with qubits as vertices and gates as edges (1-qubit gates are self-loops on a single vertex). The computational graph of candidate gates can conflict by sharing a qubit, in which case these gates cannot be scheduled simultaneously. The CLS then finds the maximal cardinality matching of $G_c$ to resolve the conflicts. Figure \ref{schedule} illustrates an example of the maximal matching process.  Algorithm \ref{cls} describes the CLS process.
\begin{algorithm}[h]
 \begin{algorithmic}
     \STATE\textbf{Input:}quantum GDG $G_q$, the list of commutation groups on qubits $\{com\_list[q_i]\ |\ q_i\in $ all qubits $\}$.
 \STATE\textbf{Output:} logical schedule $S$.
 \STATE \textbf{Initialize} current gates $cg$, next\_time\_point $np$, current commutation groups \{$com\_group[q_i]\ |\ q_i \in$ all qubits\}.

\WHILE{$cg$ not empty}
\STATE candidate gates $ng=$\{$g$ can be scheduled at $np$| $g$ in  $cg$\}
\STATE gates to be scheduled $gs=$ find\_max\_matching($ng$)
\STATE $S$+=$gs$; $cg$-=$gs$
\STATE \textbf{Update} $np$
 \FORALL{$q_i \in$ all qubits}
 \IF{$com\_group[q_i]$ empty}
 \STATE $com\_group[q_i] = com\_list[q_i].pop()$
 \ENDIF
 \ENDFOR
\STATE $cg$+=$\{g\ |g \in com\_group[q] \text{ for } q \in op(g) \}$
\STATE \textbf{Update} $com\_group$ 
\ENDWHILE
\RETURN S
 \end{algorithmic} 
 \caption{\textbf{CLS}}
 \label{cls}
\end{algorithm}

Similar to previous work \cite{Giacomo2017}, our strategy is intended to maximize parallelism, and not to minimize the number of SWAP gates in the backend. Our motivation for this strategy in our compilation scheme is the finding that SWAP gates can be beneficial in reducing latency on superconducting architectures \cite{schuch2003}, so we don't aim to reduce the amount of SWAP gates. We also believe that a precise cost model that correctly discriminates the latency of each SWAP gate in circuits leads to more efficient scheduling strategies. We propose it as an exciting open problem.

\subsection{\label{sec:backend}Compiler backend}

The backend is responsible for mapping level optimization and final pulse generation. The backend executes the following steps:  qubit mapping, topological constraint solving, and final instruction set aggregation.

\subsubsection{Qubit mapping \& topological constraint resolution:}
Our logically-scheduled instructions do not account for the topological connectivity constraints of the underlying hardware. For the benchmarks presented in this paper (Table \ref{bm}), we assume a rectangular-grid qubit topology with two-qubit operations only permitted between direct neighbors. This topology is representative of typical near-term superconducting quantum computers \cite{IBM-backends}.

To conform to this topology, the logically-scheduled instructions are processed in two steps. First, we place frequently interacting qubits near each other by bisecting the qubit interaction graph along a cut with few crossing edges, computed by the METIS graph partitioning library \cite{METIS}. As described in previous work \cite{graph-mapper-github, optimized-surface-code}, this strategy is applied recursively on the partitions, yielding a heuristic mapping that reduces the distances of CNOT operations. 

Once the initial mapping is generated, two-qubit operations between non-neighboring qubits are prepended with a sequence of SWAP rearrangements that move the control and target qubits to be adjacent.

\subsubsection{Instruction aggregation:}\label{final_ai}
 In this step, the compiler iterates with the optimal control unit to generate the circuit's final aggregated instructions. The optimal control unit optimizes each instruction individually. We describe how our instruction aggregation algorithm preserves parallelism in Section \ref{agg}.

Finally, using the CLS from Section \ref{cschedule}, the compiler schedules the circuit of aggregated instructions and sends the concatenated pulse sequences to the underlying hardware. 

\subsection{Optimal control unit}\label{ocu}
The optimal control unit in our compiler backend \cite{Nelson2017} provides optimized control pulses for each aggregated instruction. Our GPU accelerated quantum optimal control algorithm is based on automatic differentiation and the Tensorflow framework. Automatic differentiation allows users to specify advanced optimization criteria and easily incorporate them in pulse generation. These criteria include realistic experimental concerns like suppressing unwanted qubit levels, avoiding large voltage fluctuation, and most importantly, pulse latency. 

The optimal control algorithm in our unit has been validated against real hardware and used in real experimental environments \cite{Heeres2017,binomial}.

\subsection{Verification}
Our framework uses the popular QuTip \cite{qutip,qutip2} simulation backend to verify the quantum unitaries defined by the aggregated instructions and the resulting pulses generated by the optimal control unit. This verification procedure provides users confidence in the numerical accuracy of the results.

For our simulation (Section \ref{experiment}), we sample 10 aggregated instructions for each benchmark to verify that the control pulses of all instructions produce the correct unitary.
	\section{Instruction aggregation}\label{calgorithm}

This section details the two algorithms for aggregating instructions in Section \ref{ps_ai} and Section \ref{final_ai}. We first discuss the allowed action space. Then we move onto our aggregation algorithms.
 \begin{figure}[h]
      \centering
\includegraphics[scale=0.39]{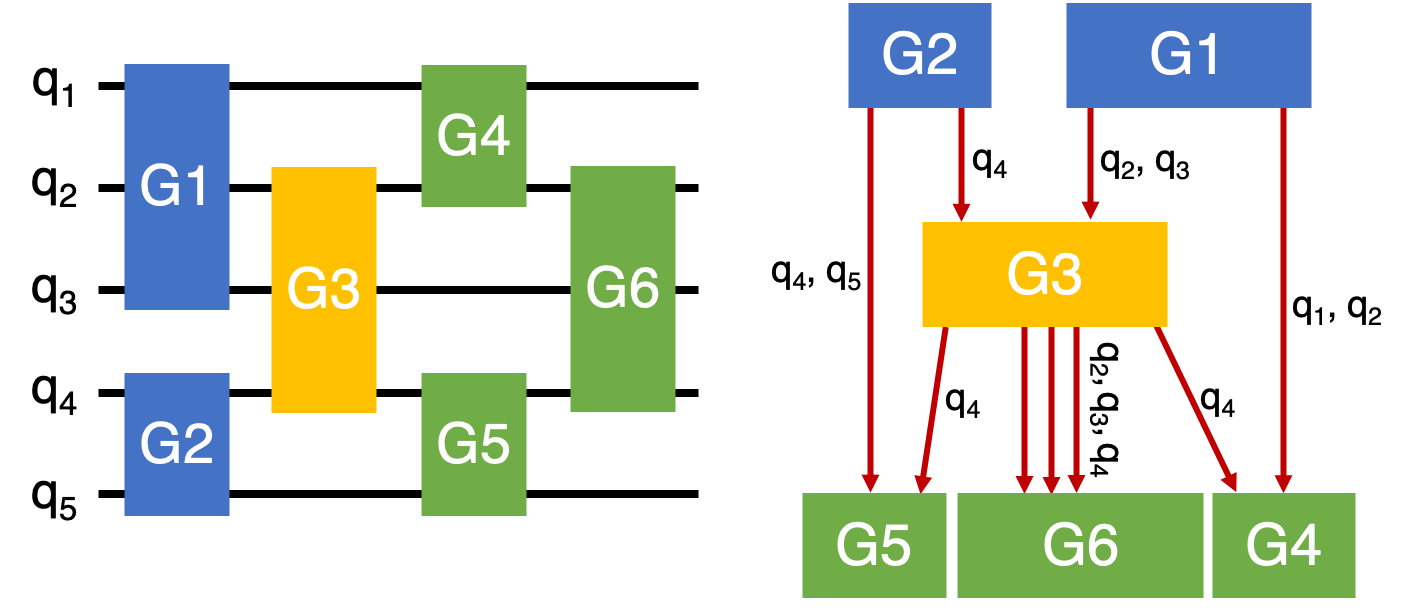}

\caption{A circuit demonstrating the action space of instruction aggregation, with the corresponding GDG to the right. Gates in the same color group commute. $G_3$ can aggregate with any of the other gates. Only the action of aggregating $G_3$ and $G_6$ is monotonic in this circuit. All other aggregation pairs induce serialization upon the circuit by delaying a dependent aggregated instruction.}
\label{algo_ps_ai}
\end{figure}

\subsection{Action space for instruction aggregation}
\label{action_space}
Here we define the allowed action space on GDG, where two instructions can aggregate if the following are true:
1. the two instructions overlap (share some common qubits); 2. one is the parent of the other on every qubit path they share or they are siblings; 3. If the two gates have parent-children relations, the parent (the children) either commutes with all gates in its commutation group on their common qubits or can be scheduled last (first) in the commutation group. In this way, we enforce the pulses inside an aggregated instruction to be continuous. In practice, we also limit the number of qubits in an aggregated instruction (instruction width) because of the scalability of the optimal control unit.
\subsection{Diagonal unitaries aggregation for commutativity detection}
To our knowledge, the most common commutative instructions are instructions representing diagonal unitaries because diagonal unitaries are used widely in decomposition methods of quantum chemistry applications \cite{qchemistry} and near-term optimization algorithms \cite{Farhi2014}. To preserve parallelism, we only detect diagonal unitaries in blocks with a width of 2 qubits. To aggregate diagonal unitaries, we exhaustively search the action space defined in Section \ref{action_space} within 2-qubit wide blocks(typically no longer than 10 gates).

\begin{figure*}[thpb]
 \vspace*{-4.2cm}
      \centering
 \hspace*{-0.3cm}\includegraphics[width=1.0\textwidth]{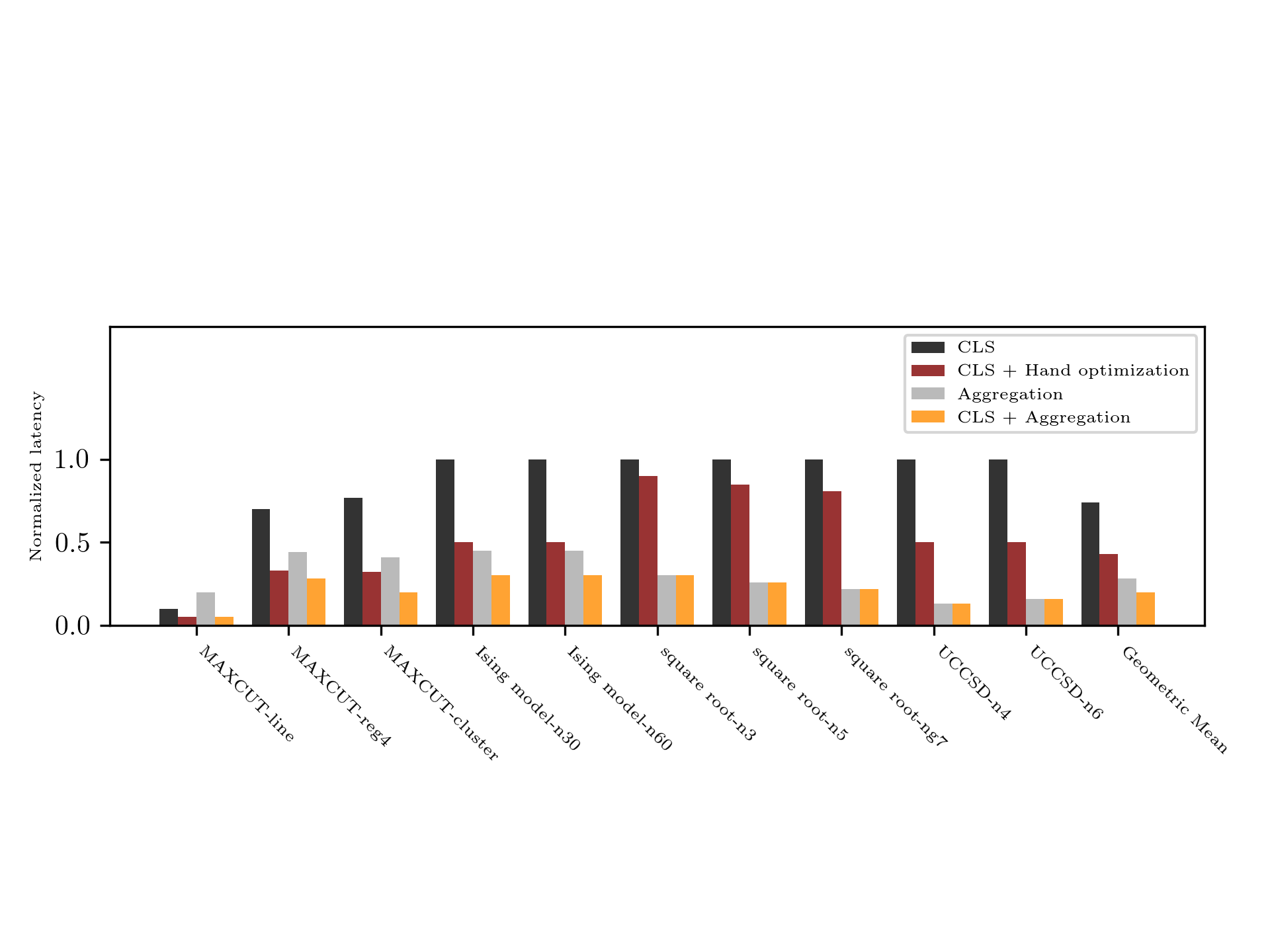}
 \vspace*{-2.4cm}
   \caption{Normalized circuit latency of different strategies (ISA compilation is the baseline with latency 1.0). } 
      \label{latency}
   \end{figure*}

\begin{table*}[]
\vspace*{0.7cm}
\begin{tabular}{llllll}
\hline
Benchmark       & Application Purpose                    & Qubits & Parallelism  & Spatial locality  & Commutativity\\ \hline

MAXCUT-line     & MAXCUT on a linear graph               & 20     & Low            & High            & High \\
MAXCUT-reg4     & MAXCUT on a random 4 regular graph     & 30     & High             & Medium            & High     \\
MAXCUT-cluster  & MAXCUT on a cluster graph     & 30     & Medium            & Low             & High   \\

Ising model     & Find ground state of Ising model       & 30     & High             & High            & Medium         \\
Ising model     & Find ground state of Ising model       & 60     & High             & High            & Medium         \\
square root-n3  & Grover algorithm for polynomial search & 17     & Low            & High          & Low     \\
square root-n4  & Grover algorithm for polynomial search & 30     & Low            & High          & Low     \\
square root-n5  & Grover algorithm for polynomial search & 47     & Low            & High          & Low     \\
UCCSD-n4           & UCCSD ansatz for VQE                   & 4      & Low                & High             & Low         \\ 
UCCSD-n6          & UCCSD ansatz for VQE                   & 6      & Low                & Medium             & Low \\        \hline
\end{tabular}
\caption{Benchmarks}\label{bm}
\end{table*}

\subsection{Instruction aggregation}\label{agg}

The main challenge of aggregating proper multi-qubit instructions is the conflict between parallelism and the need for larger instruction size for more speedup.  Aggregating new instructions may potentially compromise parallelism. For example, in Figure \ref{demo_qaoa}, if $G_5$ is merged with $G_3$, then the circuit is serialized by the delay of $G_4$, which is dependent on $G_3$.  
To protect parallelism without querying optimal control unit too often, we make the following observation: for each aggregated instruction, the larger the instruction is, the more optimized the control pulses will be. Also, we notice that there is a set of allowed actions that will not delay critical paths even if the pulses in the new instruction are not optimized. We call these actions monotonic actions because in these actions, the reward of reducing circuit latency from aggregating a collection of instructions is strictly higher than aggregating a subset of the collection, as parallelism is not compromised. Monotonic actions can be checked by explicitly calculating the original circuit depth with the depth upon executing the action. 

Our strategy is first to traverse the GDG. For each instruction in the GDG, we search the monotonic action set and keep the best action in a global table. After traversal on the GDG, we perform the global best action, and update the GDG and action table. We repeat until no more actions can be made. Then we update the latency of each aggregated instruction by querying the optimal control unit. This updated instruction latency could change the circuit structure and potentially create more monotonic actions, so we iteratively execute the above procedure until the GDG converges. For example, for the GDG in Figure \ref{gdg} (c), after one iteration of instruction aggregation, we transform it to the circuit in Figure \ref{gdg} (d). Figure \ref{algo_ps_ai} also illustrates an example of maintaining parallelism in the action space.
\section{Evaluation}\label{experiment}
In this section, we present our simulation results. We first introduce our benchmark methodology. Then we present the main result --- the latency between different compilation strategies.  We conclude by analyzing the different factors that affect the final latency, including instruction width, parallelism, commutativity, and spatial locality.

\begin{figure}[h]
      \centering

\includegraphics[width=0.25\textwidth]{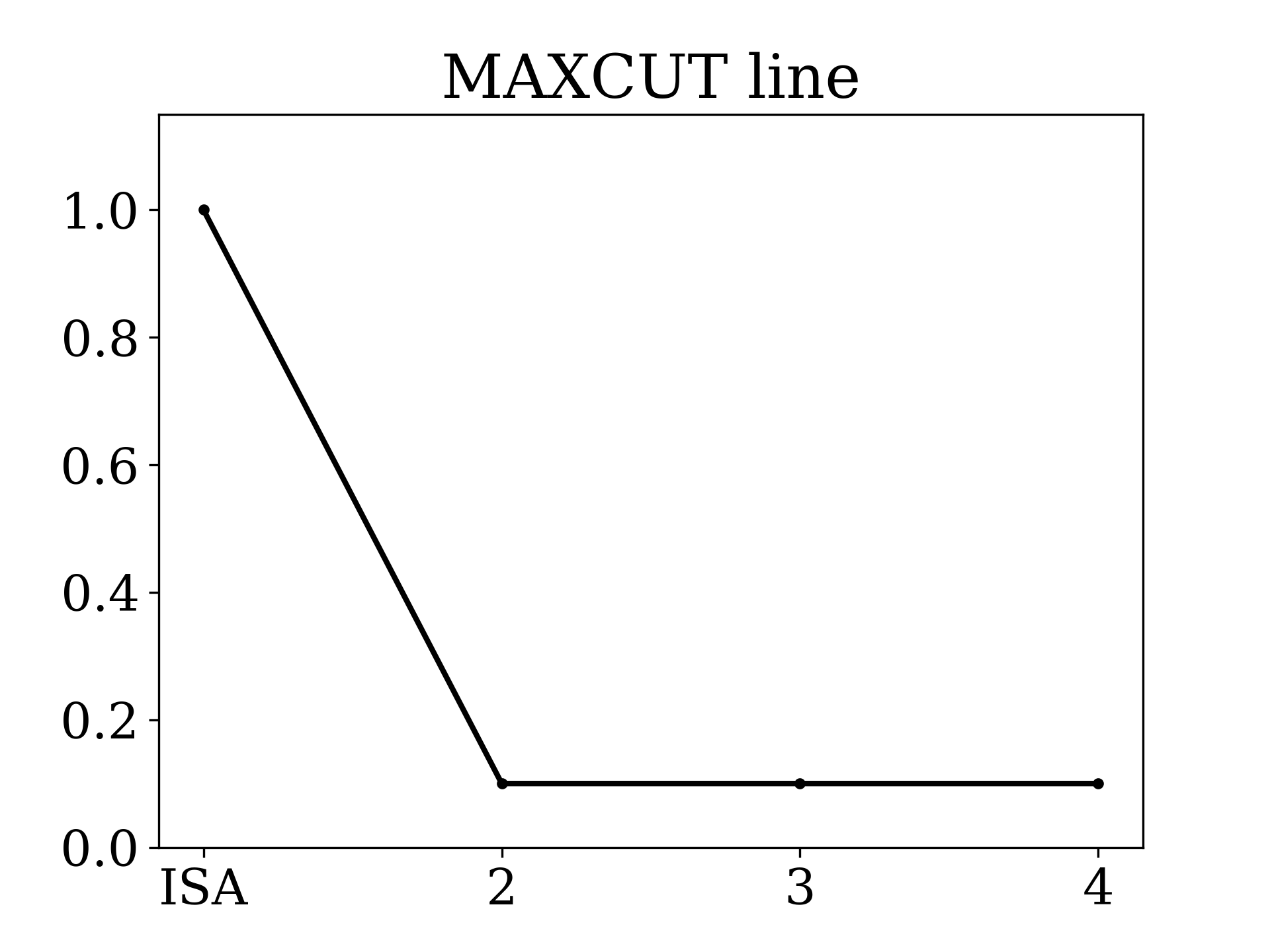}
\hspace*{-0.6cm}
\includegraphics[width=0.25\textwidth]{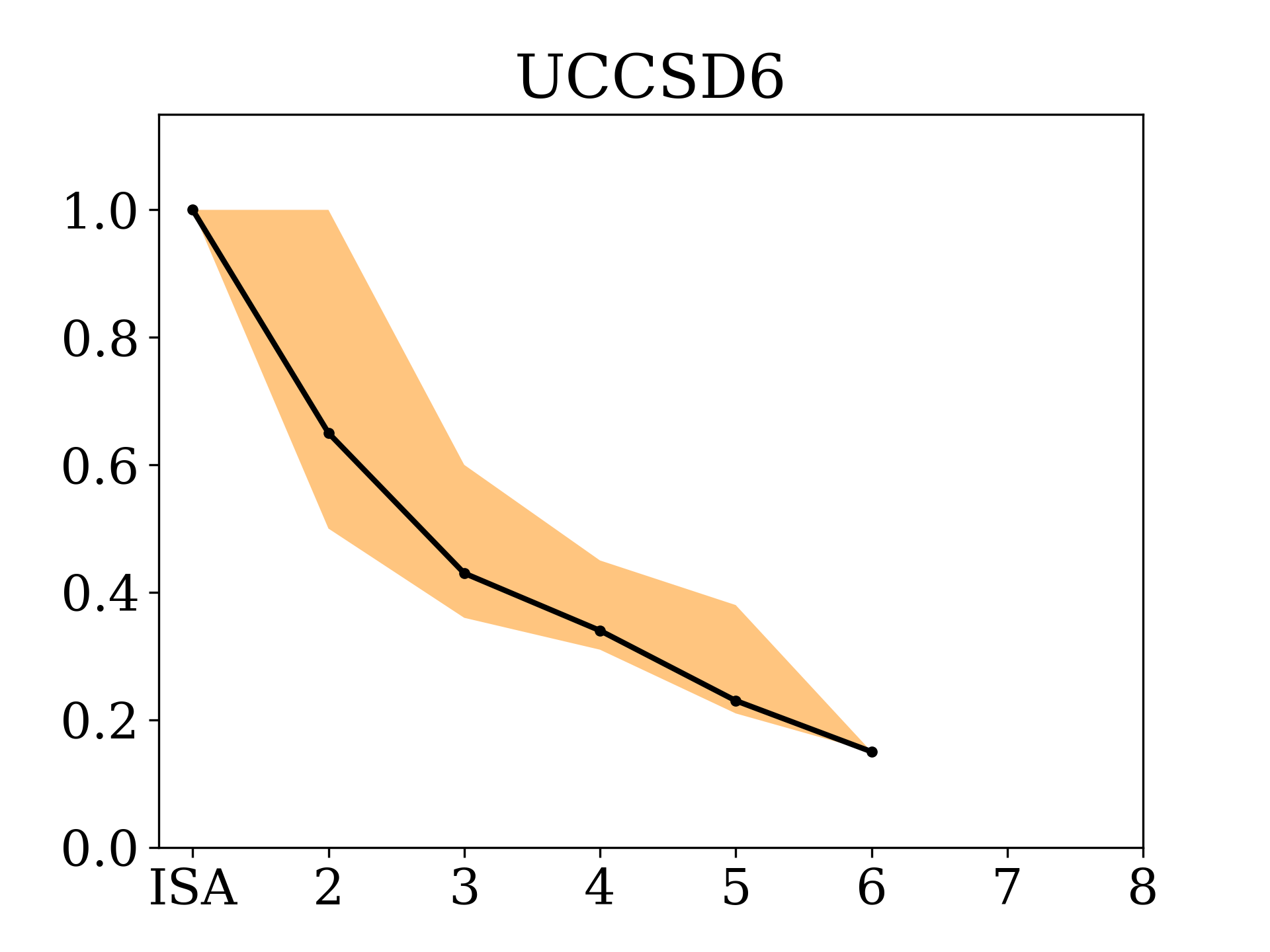}
\\
\includegraphics[width=0.25\textwidth]{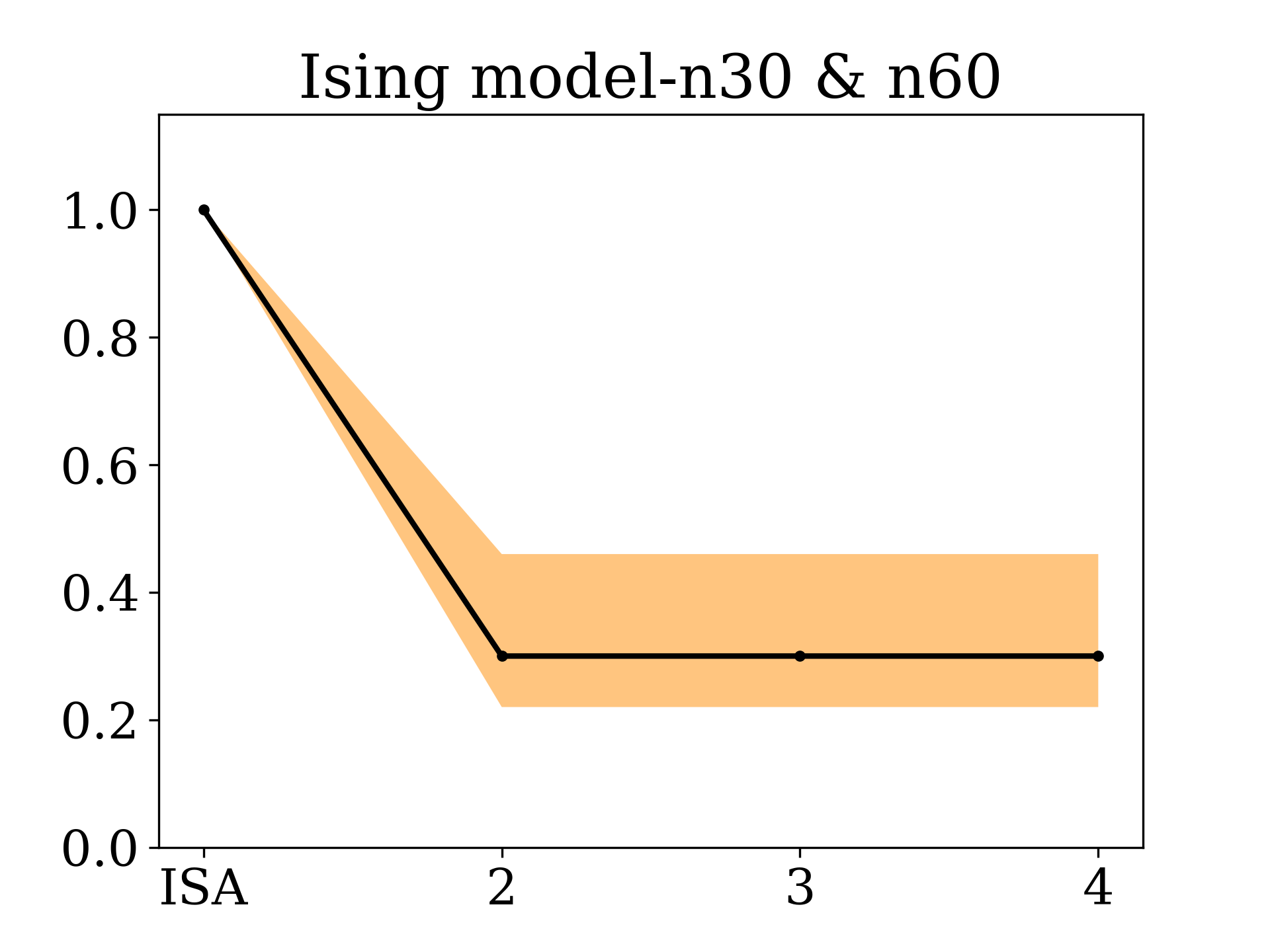}
\hspace*{-0.6cm}
\includegraphics[width=0.25\textwidth]{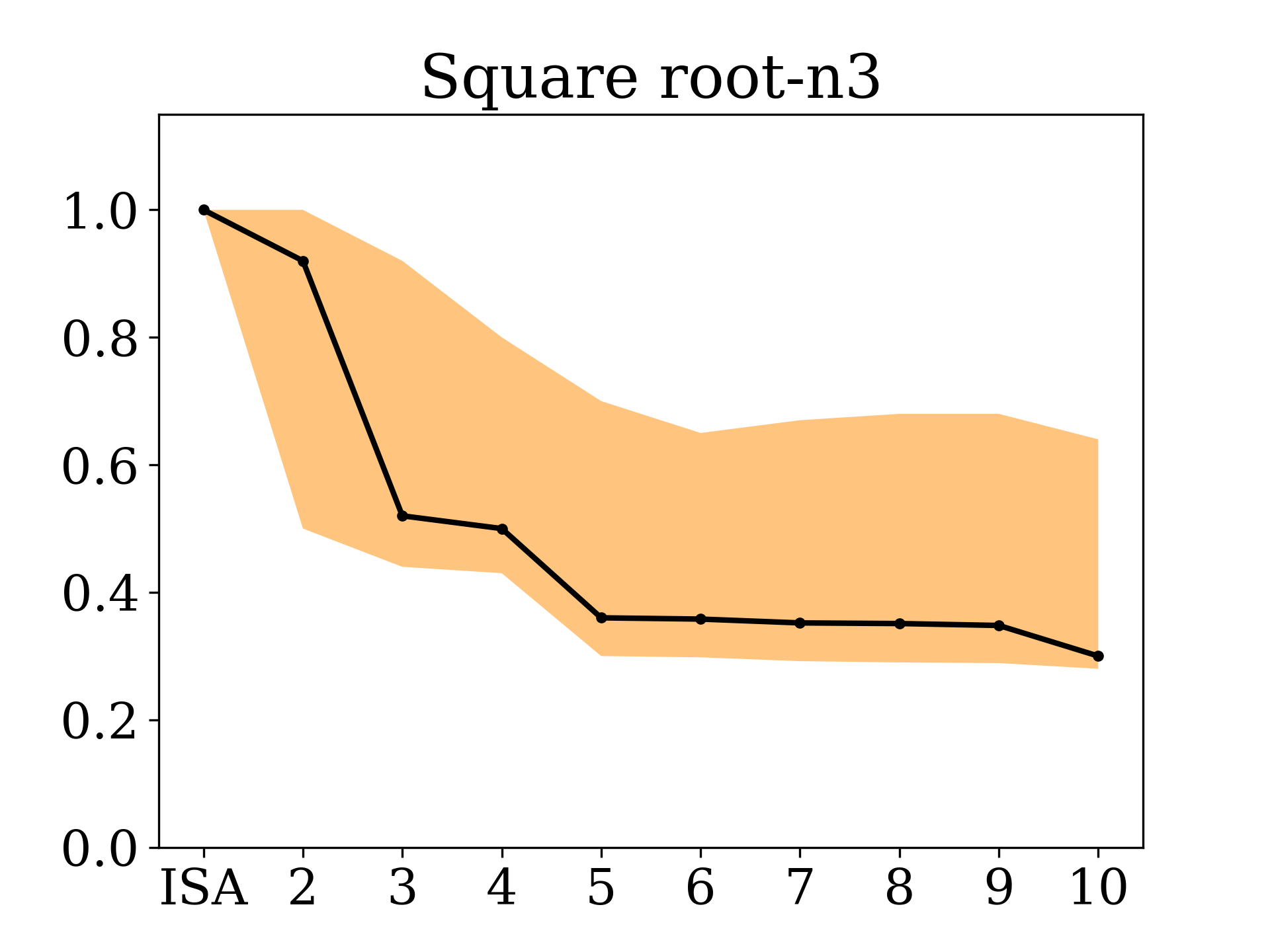}
\\
\includegraphics[width=0.25\textwidth]{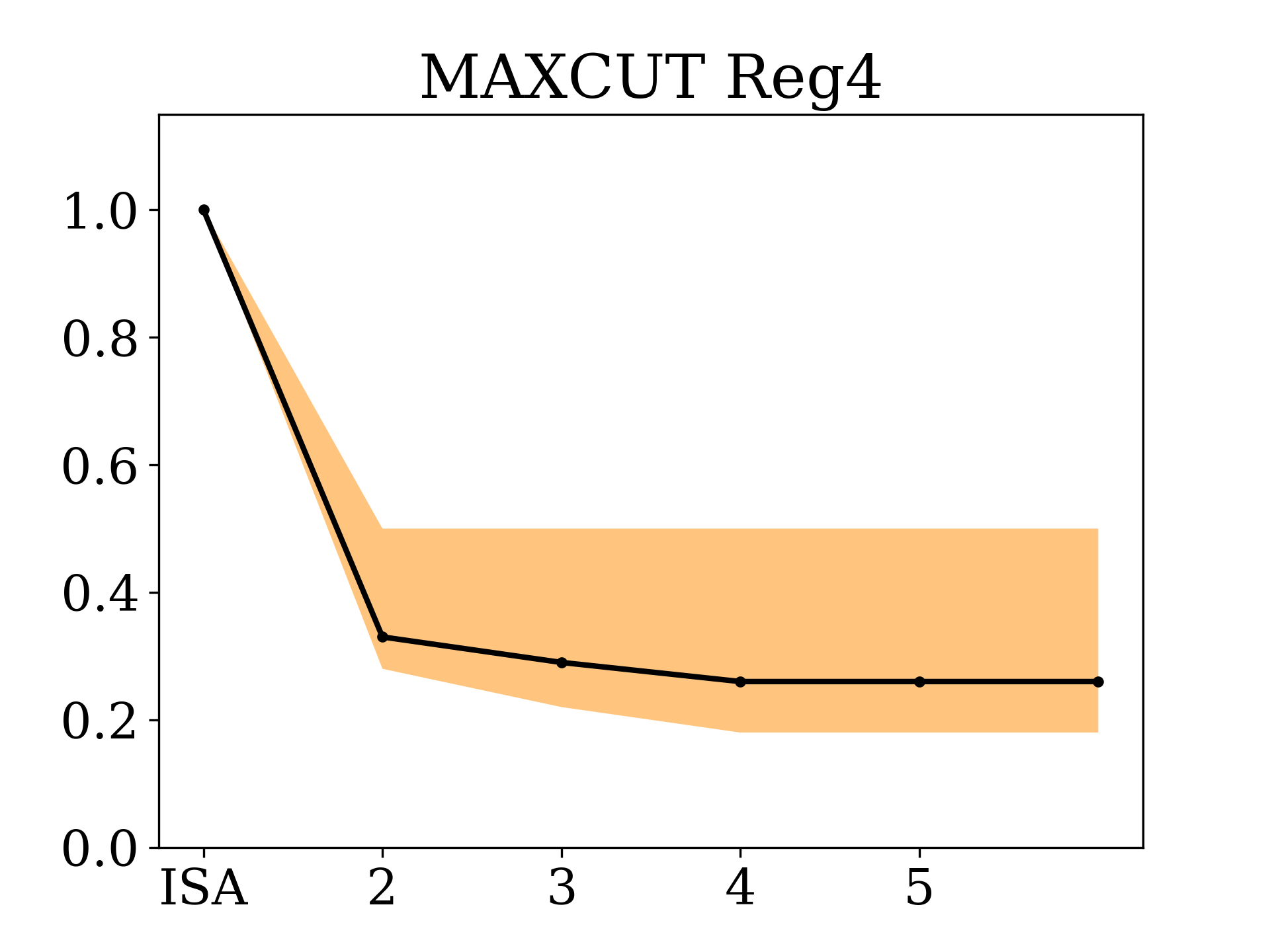}
\hspace*{-0.6cm}
\includegraphics[width=0.25\textwidth]{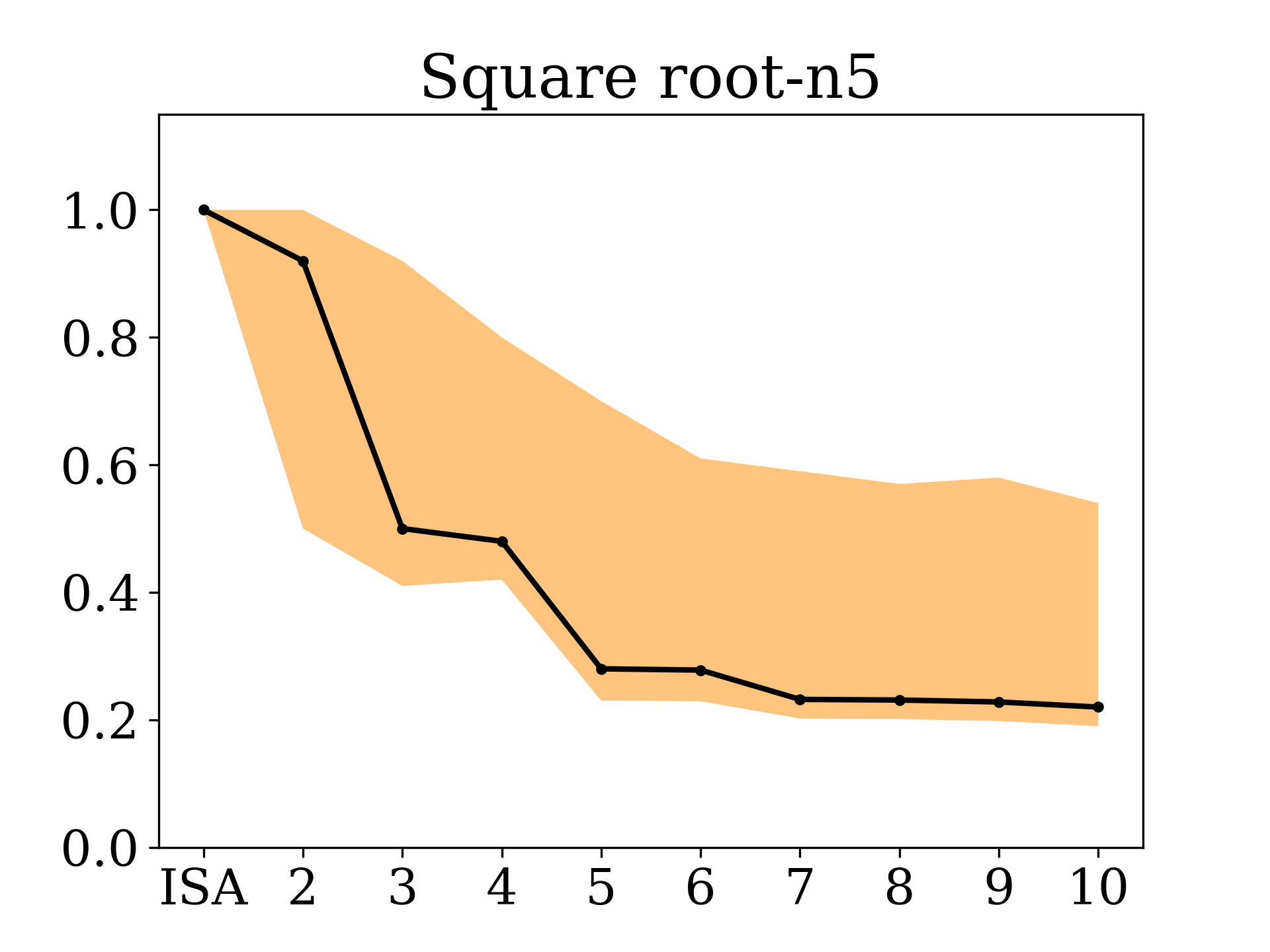}

   \caption{Allowed instruction width vs normalized latency in selected benchmarks. The black line is the normalized latency of the entire circuit. The upper (lower) edge of the filled area is the normalized latency of the instruction on the critical path that has the least (most) pulse optimization. The three applications in the left column are parallelized, either originally or after CLS. The three applications in the right column are serialized. Increasing the allowed instruction width will benefit serialized applications more. } 
      \label{para}
   \end{figure}
   \subsection{Experimental setup}
   
   We perform our numerical study on superconducting architecture with XY interaction (Appendix \ref{app1}) and set the control field limit of XY interaction to be $\mu_{max}=0.02$GHz and single qubit rotation control field to be $5\mu_{max}$.By setting the control field strength to less than typical transmon anharmonicity, we model transmon operations with low leakage to high level states \cite{jerry}.
   
\subsection{Benchmark methodology}
We select several important classical-quantum hybrid algorithms and traditional quantum applications from the NISQ era as our benchmarks. The benchmarks are chosen to have different program characteristics that will affect the improvement from the aggregated instruction compilation.
The complete list of benchmarks is shown in Table \ref{bm}. The first three benchmarks are QAOA circuits solving MAXCUT problems \cite{Farhi2014, ScaffCC}. These circuits are highly commutative. Ising model is a family of highly parallel circuits with limited commutativity \cite{ScaffCC}. Square root circuits use  Grover's algorithm \cite{grover,ScaffCC} to find the square root of polynomials and they are very serialized. UCCSD stands for Unitary Coupled Cluster Single-Double ansatz \cite{ucc} for the variational quantum eigensolver \cite{Mcclean2016}. This ansatz is derived from the Jordan-Wigner or Bravyi-Kitaev transformations \cite{qchemistry, ucc} and is considered to be a machine unaware ansatz \cite{ucc}. We include it to address  that with optimal control, physics induced ansatzs can be made more hardware efficient on superconducting architectures and more competitive relative to machine-inspired ansatzs \cite{Kandala2017}.

\subsection{Latency}
We present our main result in Figure \ref{latency}.  We compare four different strategies with unoptimized gate-based compilation. CLS refers to commutativity-aware scheduling (Section \ref{cschedule}) Aggregation 
represents executing the instruction aggregation step (Section \ref{agg}) without CLS; CLS + aggregation is self-explanatory. For hand optimization scheme, to the best of our knowledge, there are limited optimization methods documented for architectures with iSWAP gates (\cite{schuch2003,neeley2010}). Here hand optimization refers to mechanically applying the known methods (\cite{schuch2003,neeley2010}) with our best effort.

Across all 9 benchmarks, our compilation scheme achieves a geometric mean of $5.07\times$ pulse time reduction. CLS + hand optimization achieves a geometric mean of $2.338\times$ pulse time reduction. The program characteristics of each benchmark heavily affect the level of optimization by logical scheduling and aggregation, but our compilation scheme achieves better circuit latency than gate-based compilation with hand optimization for every benchmark studied here.
\section{Discussion}
\subsection{Commutativity vs Scheduling}

In our study, the level of optimization from CLS scales with the commutativity of the circuit. In applications with little to no commutativity (like square-root, QFT and UCCSD), CLS has no effect as expected. In highly commutative circuits like MAXCUT circuits, CLS alone achieves up to $5\times$ circuit length reduction. 

CLS also facilitates instruction aggregation. As shown in the Ising model-n15 example in Figure \ref{latency}, CLS alone has no optimization, but CLS + aggregation arrives at a better optimization of $3.44\times$ circuit length reduction than $2.22\times$ for aggregation alone. 

\subsection{Parallelism vs Instruction width}
 Figure \ref{para} illustrates how circuit latency reduction scales with the allowed instruction width for several applications. For highly-parallel applications such as QAOA and the Ising model, the parallelism in the circuits places limits on the instruction width of aggregated instructions. Allowing a larger instruction width, therefore does not reduce  latency. For serialized applications such as Square root and UCCSD, the latency reduction does not saturate until we reach the instruction width set by the scalability of the quantum optimal control.

In Figure \ref{para}, the lower bound of the yellow areas represents the largest latency reduction in an instruction on the critical path. In serialized applications, the total circuit latency reduction approaches this lower bound as instruction width increases. Thus, in these highly-serial applications, instructions with the largest latency reduction dominate the critical path, thus our 

\subsection{Spatial locality vs Aggregation}
To show how spatial locality affects the pulse optimization in our scheme, we compare the three instances of QAOA application in our benchmarks: MAXCUT-line, MAXCUT-reg4, MAXCUT-cluster. After CLS, all of the three instances are highly paralleled and they have similar instruction sets that are composed of the CNOT-Rz-CNOT instruction and single qubit instructions. The main difference between the three instances is the spatial locality. The less spatially localized the instance is, the more SWAP gates must be inserted in the circuit. 

 \begin{figure}[h]
      \centering
  
\includegraphics[scale=0.45]{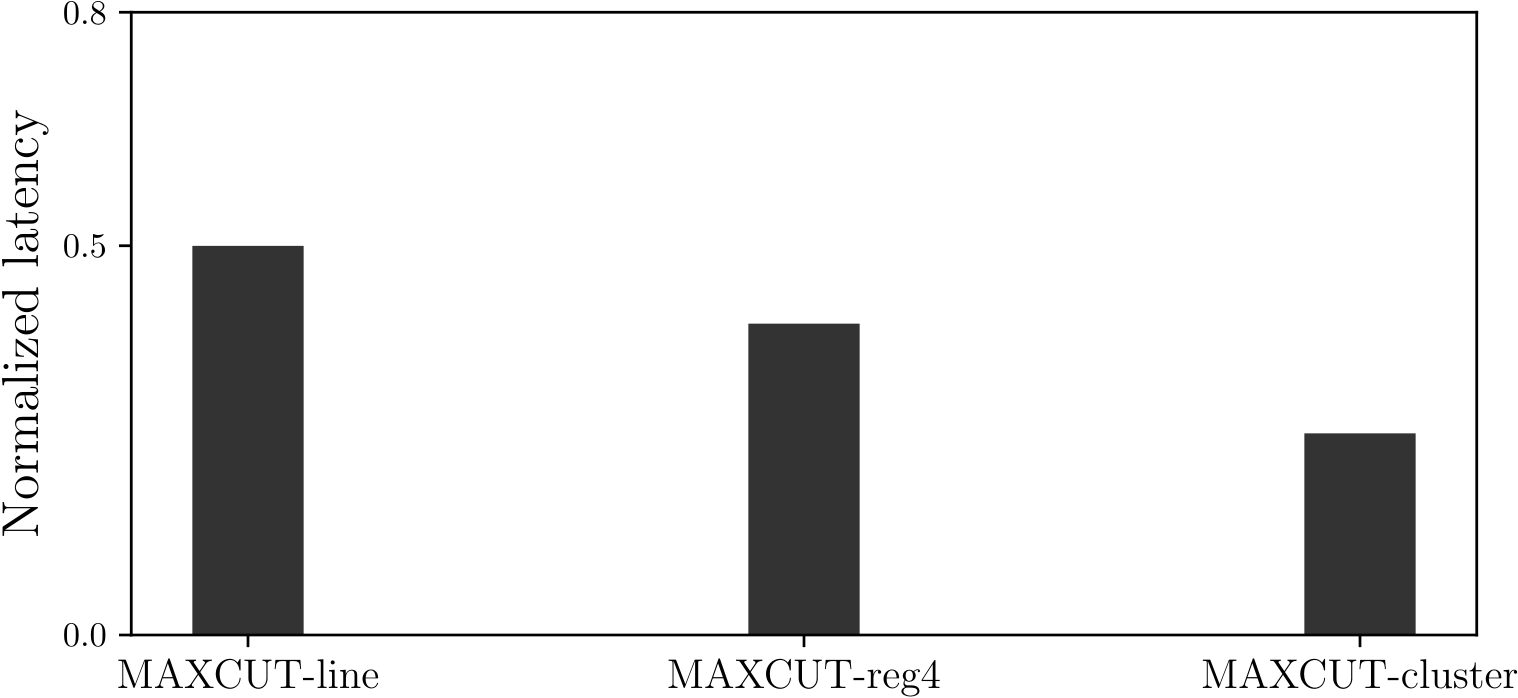}

\caption{The normalized latency of 3 instances of QAOA applications in aggregated instruction compilation scheme. For each of these instances, the latency after performing CLS is set to be 1 as baseline. From left to right, the 3 instances have high, medium, and low spatial locality. }
\label{spatial}
\end{figure}

Figure \ref{spatial} shows that the MAXCUT-cluster instance has the lowest latency and MAXCUT-line has the highest latency comparing to the normalized latency after performing CLS. For the same application, aggregated instruction compilation has larger improvements on circuits with low spatial locality.   
\subsection{Information encoding scheme vs Pulse optimization}

Information encoding schemes affect improvement due to pulse optimization. We evaluate the effect of the information encoding scheme on pulse optimization by comparing across spatially localized instances of our benchmark applications. QAOA applications encode the MAXCUT objective function directly onto the system Hamiltonian. In this simple encoding, inefficiency arises from the manual decomposition of diagonal unitaries generated by the objective Hamiltonian onto CNOT-Rz-CNOT instructions. In the spatially localized QAOA benchmark MAXCUT-line, hand optimization achieves about the same level of optimization as our compilation. UCCSD applications map molecular structures by performing the Jordan-Wigner transformation \cite{qchemistry} and then decomposing the corresponding diagonal unitaries onto CNOT-Rz-CNOT chains. In this more complicated information encoding scheme, our tool realizes $3.12\times$ greater circuit latency reduction than hand optimization in spatially localized instance UCCSD-n4. Our square root application involves reversible logic synthesis and quantum level decomposition, resulting in an encoding scheme that is more sophisticated than QAOA and UCCSD. In our Square root application, our tool realizes $3.68\times$ more circuit latency reduction than hand optimization.

From above observations, we see the trend that the more complicated the information encoding scheme is, the more advantageous our compilation is compared to hand optimization. This is expected, as simple hand optimization by replacing strategy is not efficient in finding the optimal path for complex quantum evolution with many degrees of freedom, especially when it involves sophisticated information encoding.
\section{Related work}\label{related}
Standard gate-based compilation is a well studied subject \cite{Haner2018, prac_ISA, openqasm,layered, Fu2017}. Practical techniques have been developed to improve the standard gate-based compilation from the reversible logic level down to the technology level, including studies of hand optimization, discrete \cite{Maslov_Toffoli, Maslov2008} and continuous \cite{Maslov_continuous} template matching, and rule-based rewriting \cite{whitedot, Miller2010}. Template matching methods achieved impressive gate reduction on small and intermediate-scale circuits, though they are limited by having to manually search for new template rules for each specific gate library (for example, there is no library for iSWAP gates). Rule-based rewriting methods suffer from the huge search space of rewriting strategies and  apply mainly to reversible level decomposition.

 Because the abstraction of logical level instruction remains intact in the frontend of our compilation, our workflow is compatible with most of the optimization methods described above at logical gate level. However, these upper level optimization efforts might be canceled in the backend, {\it e.g.}, if template matching takes place within an aggregated instruction, it will cause no effect because the output unitary is the same.
 
Recent work has moved beyond standard ISA abstraction. Chuang et al \cite{Chuang2016} design a new Hamiltonian simulation method that reduces the problem of quantum simulation to optimal quantum control of single qubit rotations \cite{Chuang2016}. Google proposes a plan to construct random circuits to demonstrate quantum supremacy at the pulse level  \cite{blueprint_2018}. 
   
The use of optimal control to compile large-scale quantum circuits was first explored by Schulte-Herbrueggen et al. \cite{CISC2007} in their restricted recursive-style complex quantum instructions where they report speedups up to 300\%. The researchers, however, did not provide an instruction aggregation algorithm. 
 
In this work, we provide a systematic and universal way to reduce the circuit latency the overcomes the disadvantages of previous works.
\section{Conclusion}\label{conclusion}

In this paper, we present and analyze a new compilation methodology utilizing quantum optimal control theory. This compilation aggregates multi-qubit instructions and in this way breaks the ISA abstraction in the standard gate-based compilation scheme, resulting in a competitive pulse time reduction. Our implementation of this compilation methodology shows that in several important near-term quantum applications, our compilation process achieves up to 10X circuit latency reduction on superconducting architectures, which helps enable many appealing applications. We further analyze how different program characteristics, including parallelism, commutativity, and connectivity, interact with the level of optimization by instruction aggregation. We observe that our compilation scheme is most advantageous for quantum circuits that are highly serial, have low spatial locality, and utilize sophisticated information encoding.
\section{Future work}\label{future}

There are several promising directions we propose for future study.

Compared to gate-based compilation, our scheme requires more computational resources and has a longer compilation time. For our benchmarks, the compilation time can be as long as several hours if the circuit has aggregated gates of 10 qubits. For classical-hybrid applications sensitive to long compilation time, future improvement of our compilation method is required. Partial compilation is a promising direction for solving this problem.

 Our compilation method customizes aggregated instructions for each circuit, which leads to an increase in calibrations performed in experimental settings. The conflict between amount of calibration and circuit latency can potentially be resolved by incorporating realistic error modeling into our optimal control tool \cite{optimal_control_rl}.

 Another interesting area for future work is the theoretical study on optimization of circuits on superconducting architectures with the iSWAP gate. With our numerical study, we expect to see progress in the development of new techniques for circuit transformation and application level optimization targeted for these platforms.
 
 We believe that finding a precise cost model for SWAP gates on superconducting architecture for better scheduling and mapping is an important problem. 
 
 Lastly, the instruction aggregation algorithm might be further improved by machine learning and tensor contraction techniques.
\section*{Appendix}
\begin{appendices}
      \section{Physical quantum gates}
      \label{app1}
      
Below, we list some physical gates in different architectures:
\begin{itemize}
    \item  In platforms with Heisenberg interaction Hamiltonian, such as quantum dots \cite{Kane1998}, the directly directly implementable 2-qubit physical gate is the $\sqrt{\text{SWAP}}$ gate (which implements a SWAP when applied twice).
    \item  In platforms with ZZ interaction Hamiltonian, such as superconducting systems of Josephson flux qubits \cite{paik2016,zz1999} and NMR quantum systems \cite{NMR_Chuang}, the physical gate is the CPhase gate, which is identical to the CNOT gate up to single qubit rotations.
    \item In platforms with XY interaction Hamiltonian, such as  capacitively coupled Josephson charge qubits ($e.g$ transmon qubits \cite{transmon}), the 2-qubit physical gate  is iSWAP gate.

 \item For trapped ion platforms with dipole-chain interaction, two popular physical 2-qubit gates are the geometric phase gate \cite{geophase} and the XX gate \cite{Debnath2016}. 
\end{itemize}

  \end{appendices}
  \section*{Acknowledgment}
The authors would like to thank Yao Lu, Ali Javadi Abhari, Ken Brown, James Leung, Isaac X. Shi for useful discussions. This work is funded in part by EPiQC, an NSF Expedition in Computing, under grant CCF-1730449. This work was also funded in part by NSF Phy-1818914 and a research gift from Intel. Additional funding for Henry Hoffmann comes from the DARPA BRASS program and a DoE Early Career Award.  This work was completed in part with resources provided by the University of Chicago Research Computing Center.

\bibliographystyle{plain}
 \balance
\bibliography{sys} 


\end{document}